\documentclass{appolb}
\pdfoutput=1
\usepackage{graphicx}
\usepackage[T1]{fontenc}

\def\ccqe {CC QE }
\def\numuccqe {$\nu_{\mu}n\rightarrow\mu^-p$}
\def\numubarccqe {$\overline{\nu}_{\mu}p\rightarrow\mu^+n$}
\def\numuccpip {$\nu_{\mu}N\rightarrow\mu^-N\pi^+$}
\def\numuncpi {$\nu_{\mu}N\rightarrow\nu_{\mu}N\pi^0$}
\def\ccpip {CC 1$\pi^+$ }
\def\ccpipratio {$\sigma_\textrm{\scriptsize CC1$\pi$}/\sigma_\textrm{\scriptsize CCQE}$ }
\def\ccpiplikeratio {$\sigma_\textrm{\scriptsize CC1$\pi$-like}/\sigma_\textrm{\scriptsize CCQE-like}$ }
\def\ncpi {NC 1$\pi^0$ }
\def\maqe {M$_{\mathrm{A}}^\mathrm{QE}$}
\def\enu {E$_{\nu}$}
\def\q2 {Q$^{\mathrm{2}}$}
\def\gev2 {(GeV/c)$^2$}
\def\numu {$\nu_{\mu}$}
\def\numubar {$\overline{\nu}_{\mu}$}
\def\delqp {$\Delta\theta_p$}
\def\deg {$^{\circ}$}
\def\mup {$\mu\!-\!p$ }
\def\mupi {$\mu\!-\!\pi$ }

\begin{document}

\pagestyle{plain}
\newcount\eLiNe\eLiNe=\inputlineno\advance\eLiNe by -1

\title{Measurements of Neutrino Cross Sections Near 1 GeV 
\thanks{Presented at the 45th Winter School in Theoretical Physics ``Neutrino Interactions: from Theory to Monte Carlo Simulations'', L\k{a}dek-Zdr\'oj, Poland, February 2--11, 2009.}}
\author{M.O. Wascko
\address{Department of Physics\\
Imperial College London\\
Prince Consort Road\\
London SW7 2AZ, United Kingdom}}
\maketitle

\begin{abstract}

We summarise recent neutrino and antineutrino measurements near 1~GeV
by the K2K, MiniBooNE and SciBooNE collaborations.  We focus on
experimental methods and note discrepancies between the most commonly
used models for neutrino-nucleus interactions and recent high
statistics observations of charged-current quasi-elastic scattering as
well as charged and neutral current pion production on carbon and
oxygen. We discuss possible directions for future measurements.

\end{abstract}



\section{Introduction}

Neutrino physics is entering a new era of precision.  The need for
more precise neutrino cross section measurements in the 1~GeV region
by the next generation of oscillation experiments has been well
described~\cite{itow-debbie}. We will not discuss in detail the
effects of systematic uncertainties on the next generation of
oscillation experiments, but rather focus on recent measurements that
have exposed the shortcomings of the current theoretical models
describing neutrino-nucleus interactions.  We will primarily cover
results from the K2K, MiniBooNE and SciBooNE experiments which had
been released or presented in public conferences prior to the time of
the L\k{a}dek School (February, 2009).

In Section~\ref{sec:prev_meas}, we discuss the past measurements of
neutrino interactions near 1 GeV; in Section~\ref{sec:expts} we
describe the experiments whose data are shown in later sections; in
Section~\ref{sec:ccqe} we discuss the charged current quasi-elastic
(\ccqe) process and recent measurements of it; Section~\ref{sec:cc1pi}
covers charged-current single pion (\ccpip) production processes;
Section~\ref{sec:ncpi0} covers neutral-current single pion (\ncpi)
processes; Section~\ref{sec:nubar} covers antineutrino measurements
and in Section~\ref{sec:summary} we summarise and discuss future
directions.


\subsection{Previous Measurements}
\label{sec:prev_meas}

\begin{figure}
\center
{\includegraphics[width=2.3in]{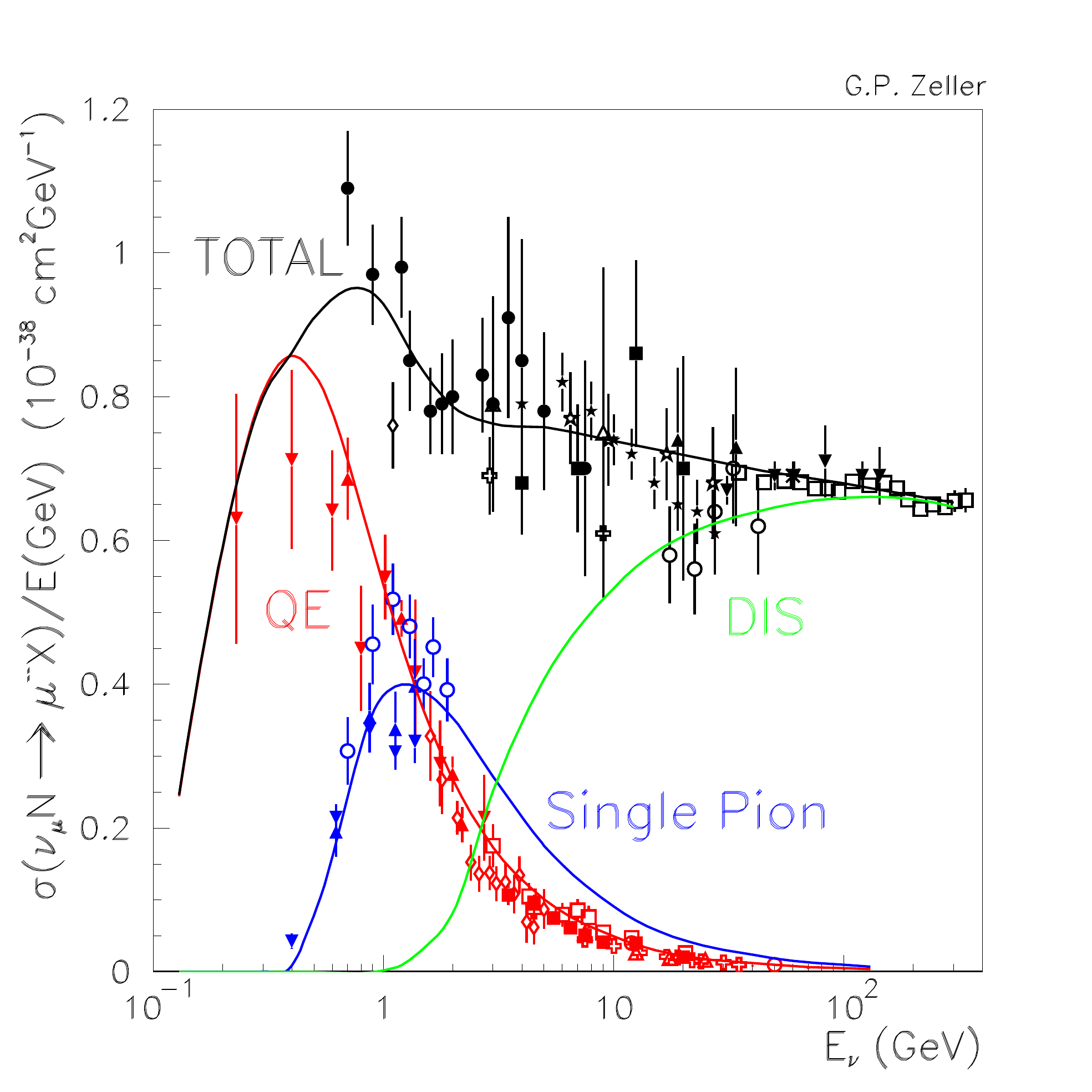}
\includegraphics[width=2.3in]{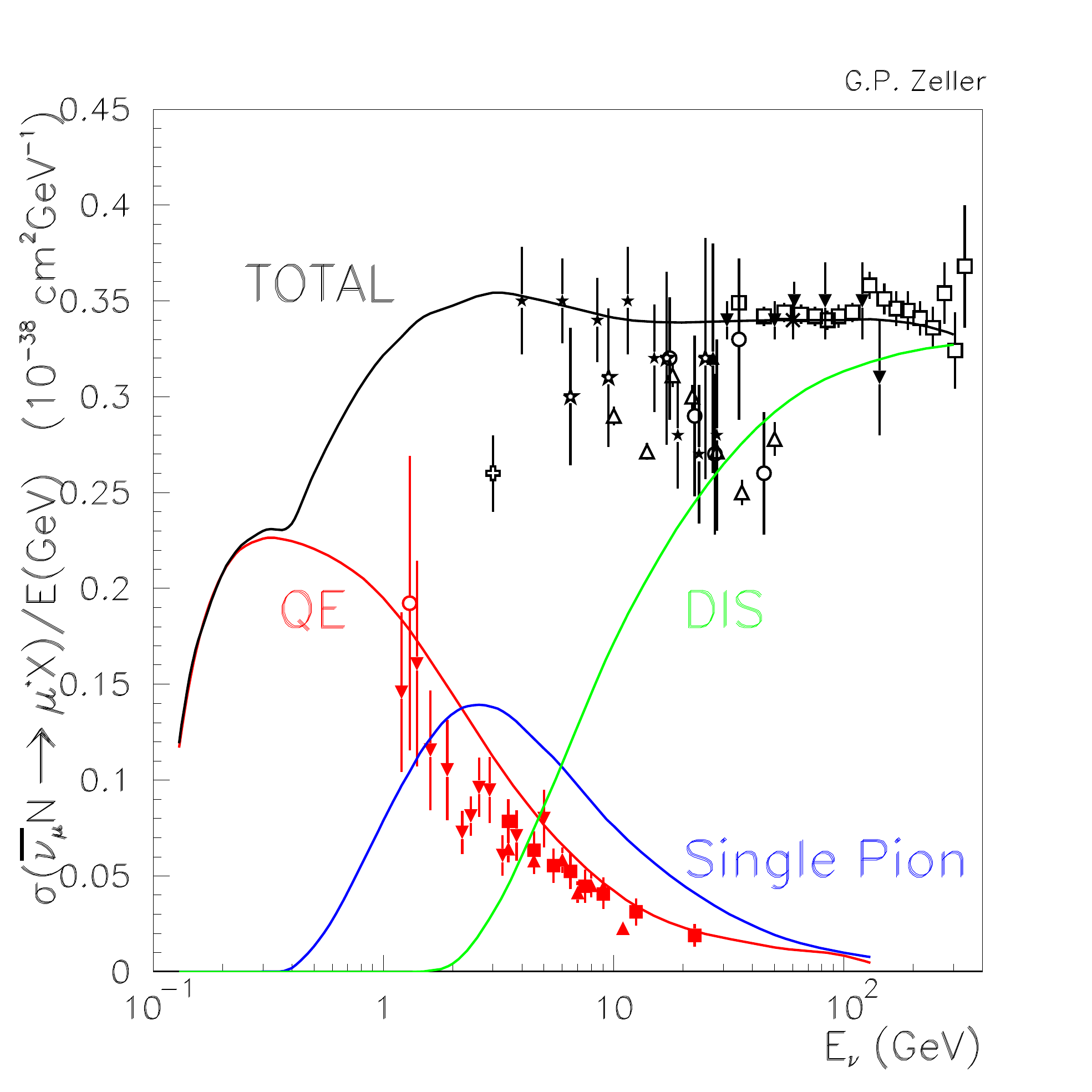}}
\caption{\em Past measurements of neutrino (left) and antineutrino
  (right) cross sections.\cite{past_data}}
\label{fig:past}
\end{figure}

Most previous measurements of neutrino interaction cross sections at
these energies were made with bubble chambers exposed to accelerator
neutrino beams; the notable exception being the Serpukhov spark
chamber.  Figure ~\ref{fig:past} summarises the past charged-current
measurements for both neutrino- and antineutrino-nucleus interactions
over a wide range of energies~\cite{past_data}.  Bubble chambers offer
extremely good final state particle reconstruction resolution, which
makes detector systematic uncertainties negligible compared to other
sources.  However, because events were necessarily reconstructed by
hand, all bubble chamber neutrino experiments collected very poor
statistics.  Lower intensity neutrino beams also contributed to lower
statistics.

\section{The Modern Experiments}
\label{sec:expts}

\subsection{Neutrino Beam Flux Predictions}

Neutrino cross section measurements require estimates of the
neutrino fluxes; these estimates have proven to be extremely difficult
since the advent of accelerator neutrino beams.  Most previous
experiments perform some calculations of neutrino fluxes based on
estimates of the secondary pion spectra; these estimates in the past
have had extremely high uncertainties.  Because of this, most
experiments employed a circular bootstrapping method of estimating the
fluxes.

\begin{figure}
\center
{\includegraphics[width=3.6in]{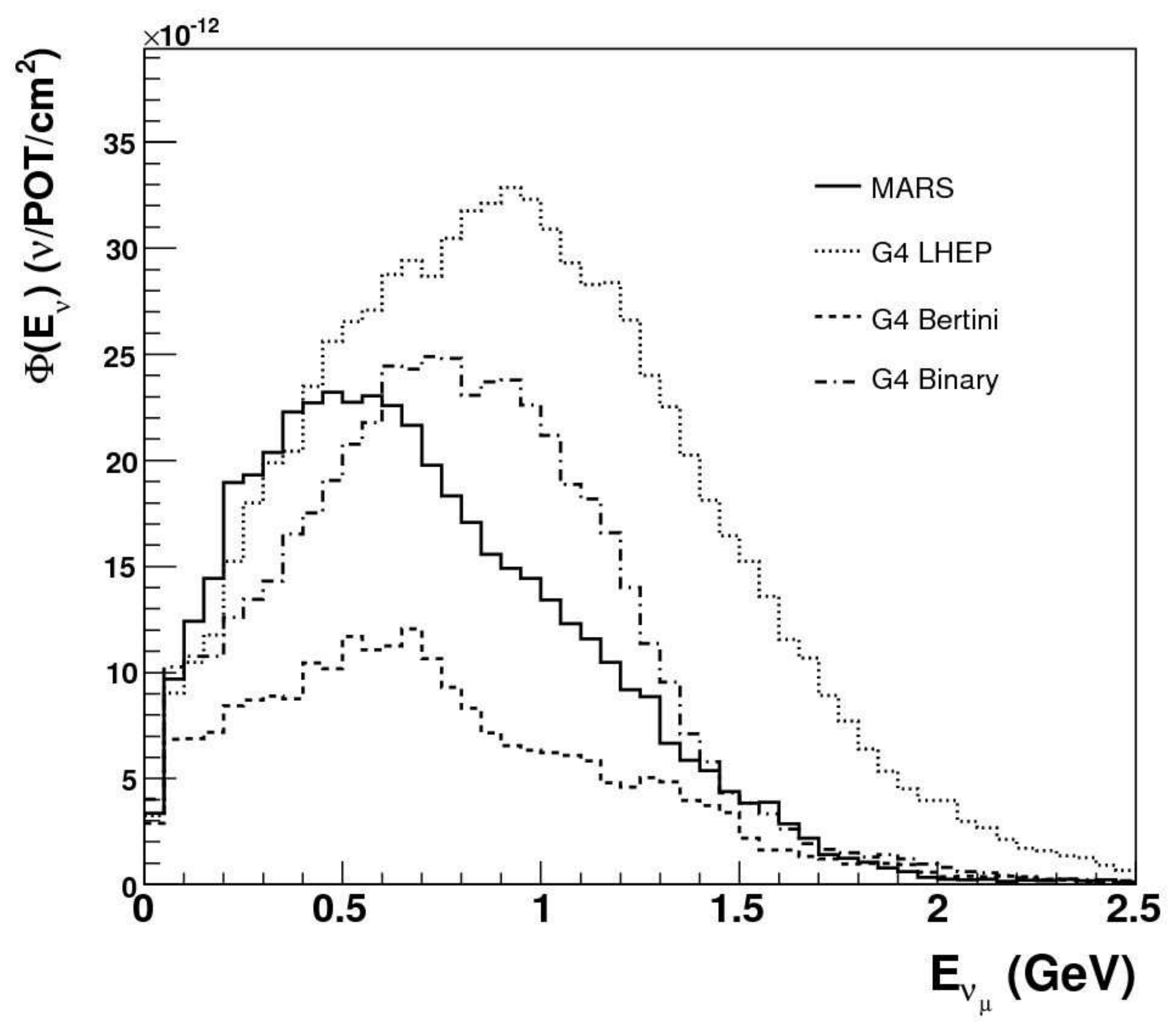}}
\caption{\em Four estimates of the neutrino flux at MiniBooNE, using
  different models for the production of parent-pions by p-Be
  collisions in the neutrino target.\cite{dschmitz}}
\label{fig:hadron_flux}
\end{figure}

To illustrate the difficulty os estimating neutrino fluxes,
figure~\ref{fig:hadron_flux} shows four examples of predicted neutrino
flux spectra at the MiniBooNE detector~\cite{dschmitz}.  Each flux
prediction was produced using exactly the same Monte Carlo (MC)
simulation of the neutrino target, horn, and secondary beamline, with
the only difference being the primary pion production in each.  The
largest flux estimate is a factor of four higher than the lowest,
illustrating in rather dramatic fashion the difficulty in estimating
neutrino fluxes.

Because of the importance of accurate neutrino flux predictions for
precise cross section measurements, several experiments have been
performed and planned to make accurate measurements of primary hadron
production cross sections.  The HARP experiment at CERN~\cite{harp}
has published precise ($\frac{\delta \sigma}{\sigma}\sim5\%$)
measurements of pion production on an aluminium target at 12~GeV for
K2K~\cite{harp_al}, and pion production on a beryllium target at 8~GeV
for the Booster Neutrino Beamline (BNB)~\cite{harp_be}, as well as
others. By explicitly measuring the production of the mesons that
contribute to neutrino production, the HARP data solve the problem
illustrated in figure~\ref{fig:hadron_flux}.  Flux predictions using
the HARP data as input have been used in publications by the K2K,
MiniBooNE and SciBooNE collaborations.  The MIPP experiment at
Fermilab~\cite{mipp} and the NA61/SHINE experiment at CERN~\cite{shine}
have also collected data which should improve the flux predictions
for the NuMI beam (MIPP) and T2K (SHINE).

\subsection{Neutrino Experiments}

K2K was a long baseline experiment in Japan which confirmed the
atmospheric oscillation signal~\cite{k2k_osc}.  MiniBooNE is a short
baseline experiment at Fermilab which successfully ruled out the
oscillation hypothesis of the LSND signal in neutrinos~\cite{mb_nue}
and is now pursuing a high statistics analysis of antineutrino data
after a preliminary result which showed no evidence of LSND-like
oscillations~\cite{mb_nuebar}.  SciBooNE is a short baseline
experiment at Fermilab designed to make precise neutrino and
antineutrino cross section measurements on carbon~\cite{sb_proposal}.

K2K is comprised of three main pieces, an accelerator
neutrino beam, a near detector suite and a far detector~\cite{k2k_osc}.
The neutrino beam is produced by impinging 12~GeV kinetic energy
protons on an aluminium target.  Secondary charged pions and kaons are
bent forward by two toroidal magnetic focussing horns into a 200~m
decay volume; the horns increase the neutrino flux in the detector
halls by a factor $\sim$22.  The result is a beam with mean neutrino
energy near 1.3~GeV.  The neutrinos first pass through the near
detectors located 300 m downstream of the neutrino target.  The first
near detector is a one kiloton water Cherenkov detector (1kT) and the
second is a fine-grained detector (FGD) which is comprised of several
subsystems: a scintillating fibre detector (SciFi), a lead glass
calorimeter (LG), a totally active scintillating bar detector (SciBar)
and a muon range detector (MRD).  The far detector used for the
oscillation analyses is the 50 kiloton water Cherenkov Super
Kamiokande detector.  K2K took data from June 1999 until November 2004.

MiniBooNE consists of an accelerator neutrino beam and a mineral oil
Cherenkov detector.  The Booster Neutrino Beam (BNB), which feeds
neutrinos to MiniBooNE and SciBooNE, is produced by impinging 8~GeV
kinetic energy protons on a beryllium target.  The secondary pions and
kaons are focused by a single magnetic horn which increases the
neutrino flux by a factor $\sim$7.  The resultant neutrino beam has
a mean energy of 0.8~GeV~\cite{mb_flux}.  The polarity of the magnetic
horn can be reversed, producing an antineutrino beam with mean energy
$\sim$0.6~GeV. The MiniBooNE detector is a 0.8 kiloton mineral oil
Cherenkov detector located 541~m from the neutrino
target~\cite{mb_nim}.  MiniBooNE began collecting beam data in
February 2002 and is approved to continue collecting data through at
least 2010.

The SciBooNE experiment is a new detector placed in the BNB at 100~m
from the neutrino target~\cite{sb_proposal}.  The neutrino vertex
detector is SciBar, the same fully active scintillating bar detector
used in K2K.  SciBooNE also uses an electromagnetic calorimeter (EC)
placed immediately downstream of SciBar and an MRD that is different
from the one used in K2K.  SciBooNE took neutrino beam data from June
2007 until August 2008.

Several more experiments have already been running or will be coming
online in the near future to make precise neutrino cross section
measurements.  Foremost among these is the MINER$\nu$A experiment at
Fermilab which started commissioning with neutrino data in the NuMI
beamline beginning in April 2009~\cite{minerva}.  MINER$\nu$A will have
the flexibility of the NuMI beamline allowing it to measure neutrino
and antineutrino cross sections over a wide range of energies, as well
as the capability to change nuclear targets. The MINOS experiment has
been running for years in Fermilab's NuMI beamline~\cite{minos}.  Also
beginning to take data in 2009 in Fermilab's NuMI beam is
Argoneut~\cite{argoneut} a liquid argon time projection chamber
(LArTPC) which will make new measurements of neutrino cross sections
on argon.  The T2K near detectors will be up and running in 2010
affording high statistics neutrino cross section
measuremnts~\cite{t2k_nuint07}, and the NO$\nu$A near detector will make
several cross section measurements~\cite{nova}.  Finally beginning in
2012 the MicroBooNE experiment, a large LArTPC, will begin running in
Fermilab's BNB~\cite{microboone}.


\section{Charged Current Quasi Elastic Scattering}
\label{sec:ccqe}

The \ccqe process, \numuccqe, is important because it is the signal
reaction for oscillation experiments in the 1~GeV region.  It is used
as the signal process because it is the largest neutrino-nucleus cross
section below $\sim$2~GeV and because the simple final state allows
accurate neutrino energy reconstruction using only the measured energy
and angle of the outgoing lepton.

The neutrino-nucleon \ccqe  scattering cross section is most
commonly written according to the Llewellyn-Smith
prescription~\cite{llewellyn-smith}, which parameterises the cross
section in terms of several form factors which are functions of \q2
$\!$, the square of the four-momentum transferred to the nuclear
system.  Many of the form factors can be taken from electron
scattering experiments.  However, the axial form factor can only be
meausred by neutrino scattering.  In the past, most experiments have
assumed a dipole form for the axial form factor $F_A$, $F_A(Q^2)=
F_A(Q^2=0)/(1+Q^2/M_A^{QE})^2)^2$, and used reconstructed \q2 \,
distributions to extract a value for the axial mass parameter \maqe.

To approximate the nuclear environment, the relativistic Fermi gas
(RFG) model of Smith and Moniz is used by most
experiments~\cite{smith-moniz}.  This model assumes that nucleons are
quasi-free, with an average binding energy and Fermi momentum which
are both specific for particular nuclei.  Pauli blocking is included
in the model.  While such simple models have been demonstrated
inadequate for electron scattering experiments,
previous neutrino scattering measurements were not sufficicient to
demonstrate model deficiencies.

More details of the theory of neutrino-nucleus scattering are discussed
in detail elsewhere in these proceedings~\cite{udias}.


\begin{figure}
\center
{
\includegraphics[width=1.6in,height=1.6in]{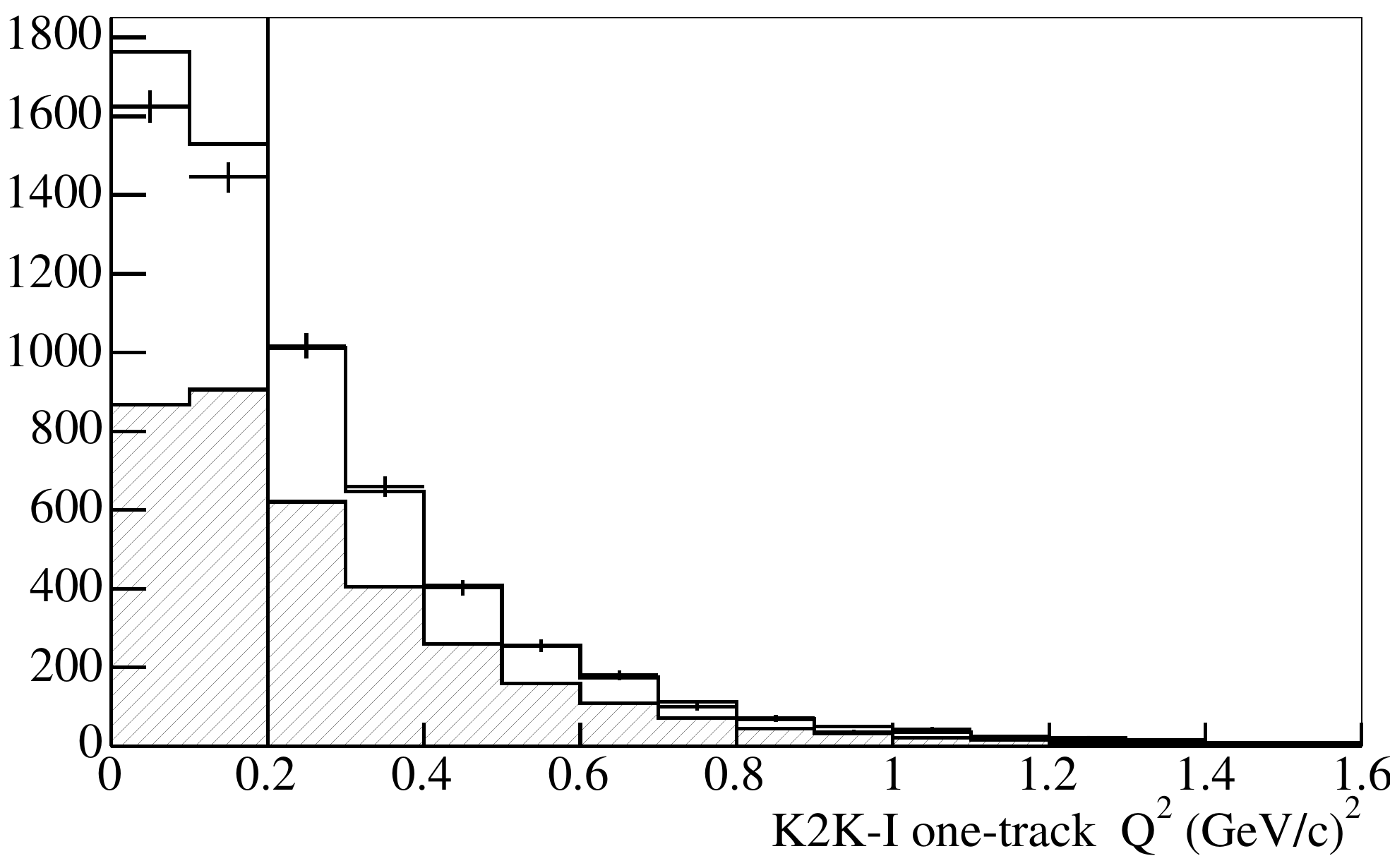}
\includegraphics[width=1.6in,height=1.6in]{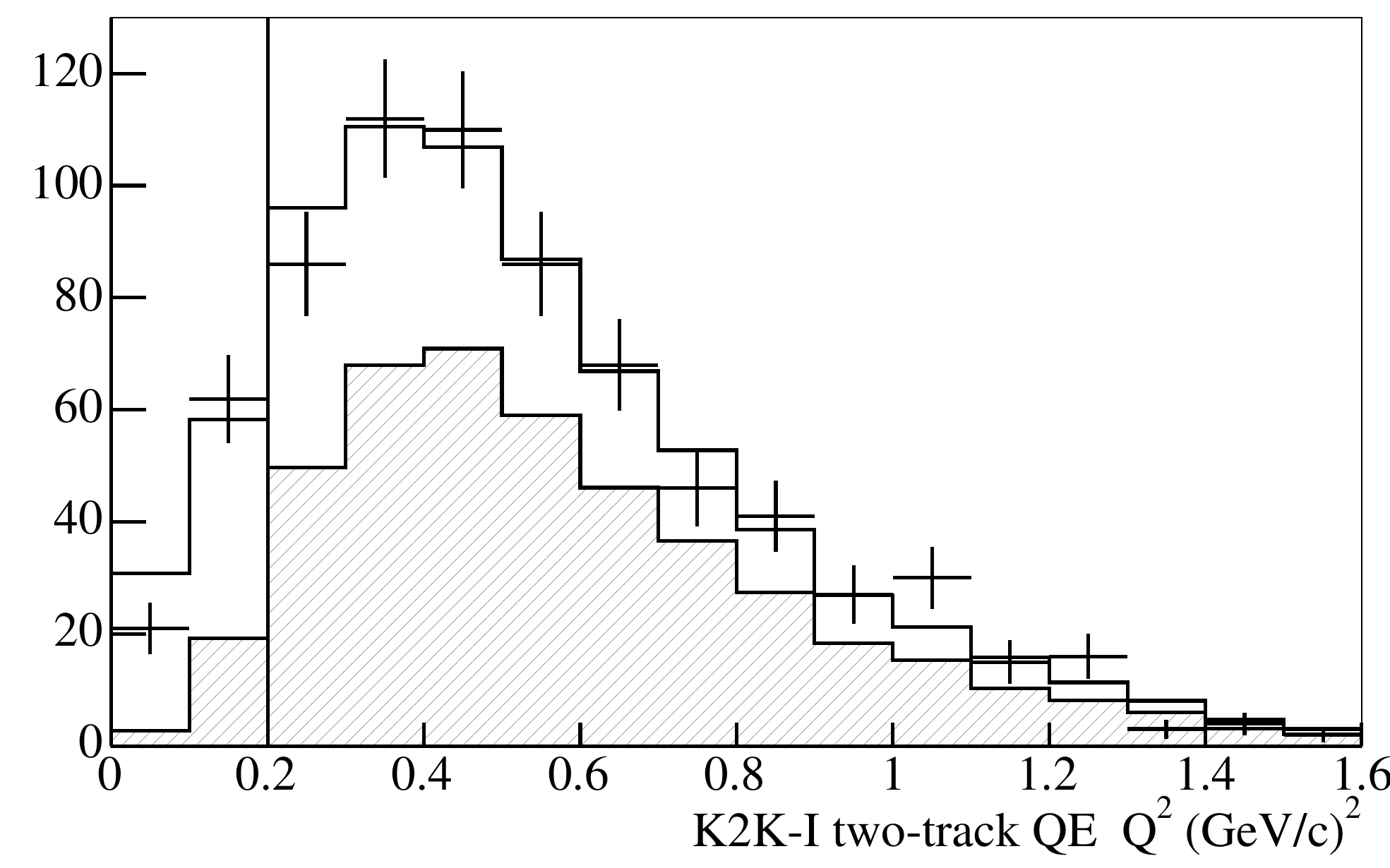}
\includegraphics[width=1.6in,height=1.6in]{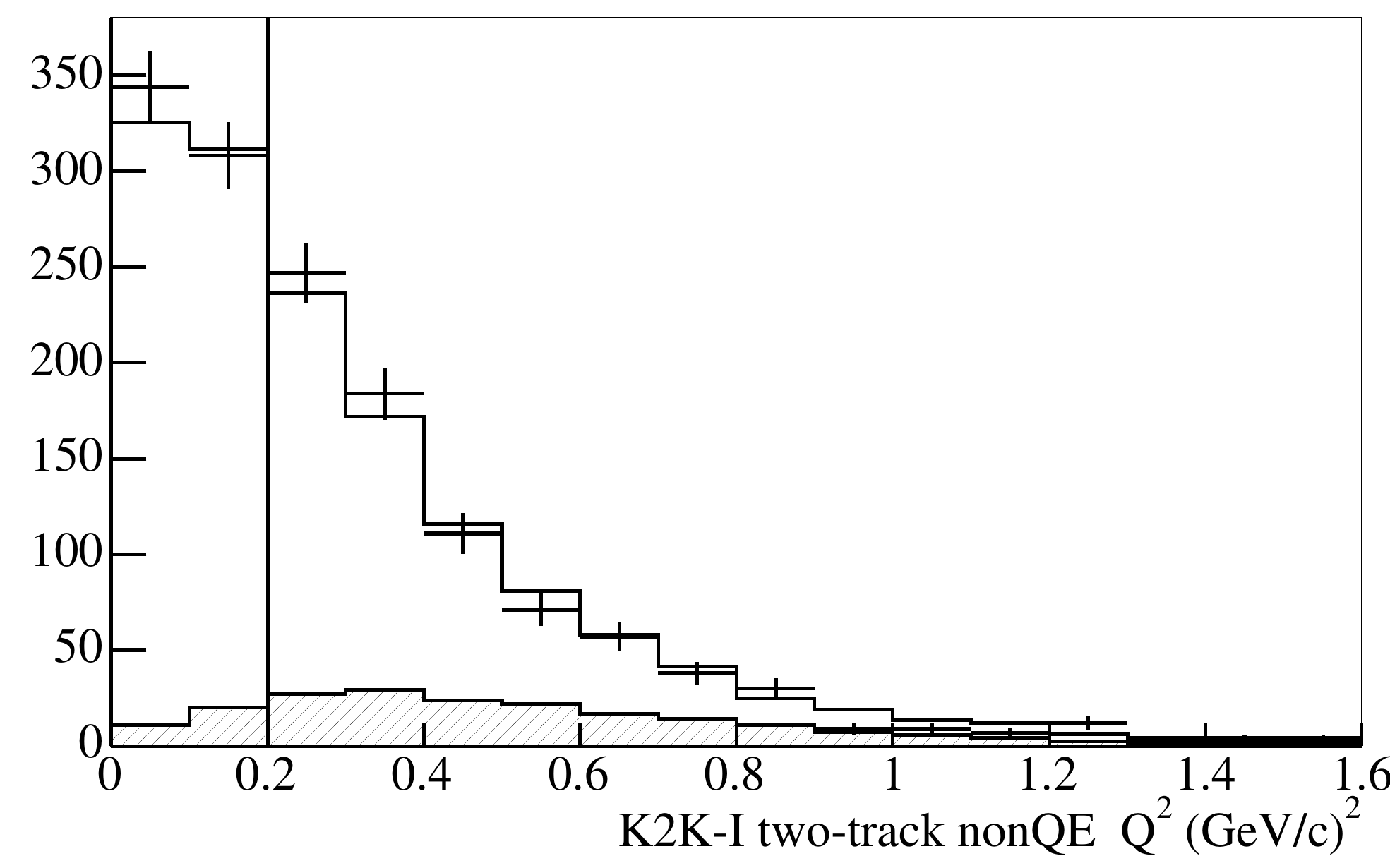}
\includegraphics[width=1.6in,height=1.6in]{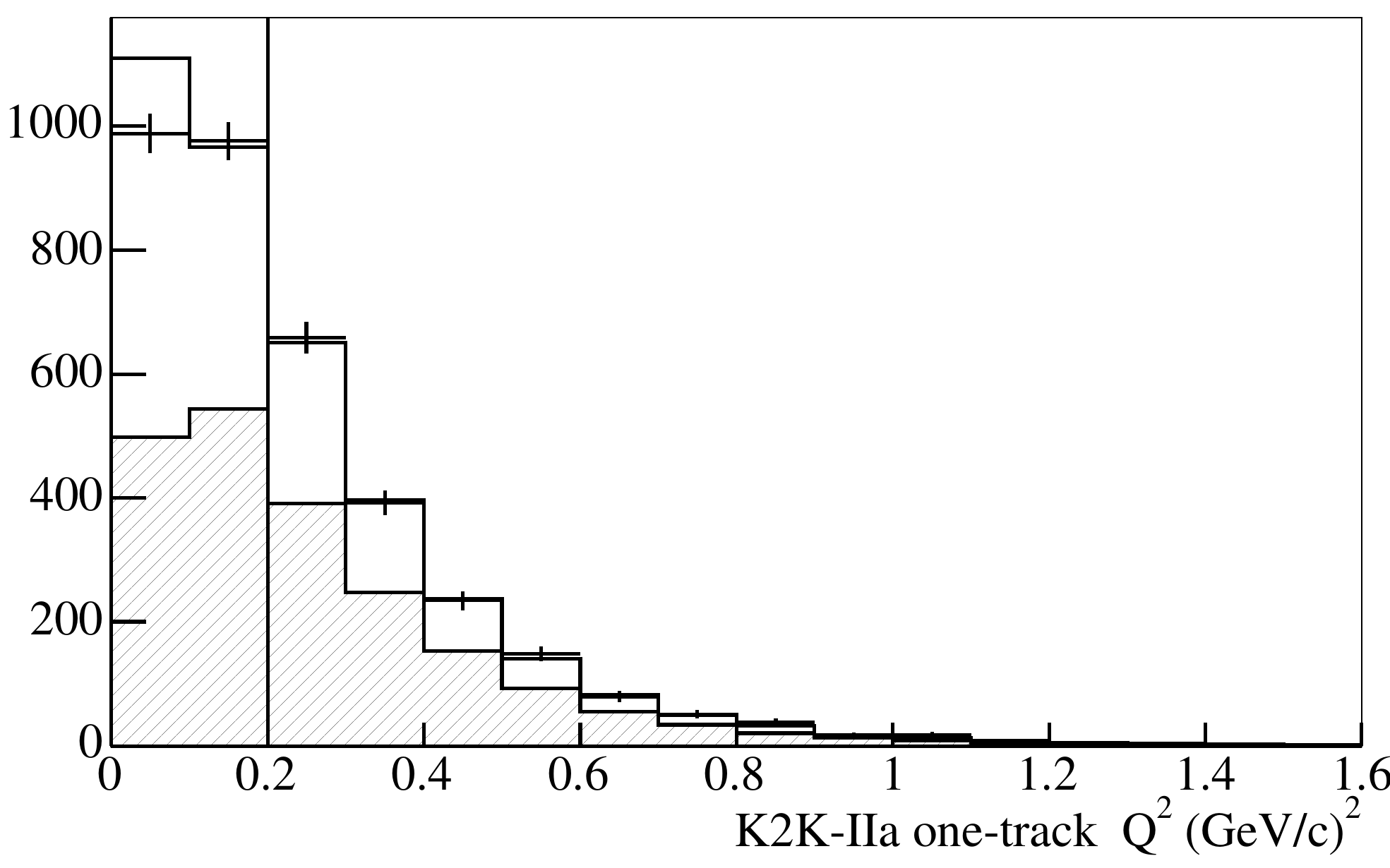}
\includegraphics[width=1.6in,height=1.6in]{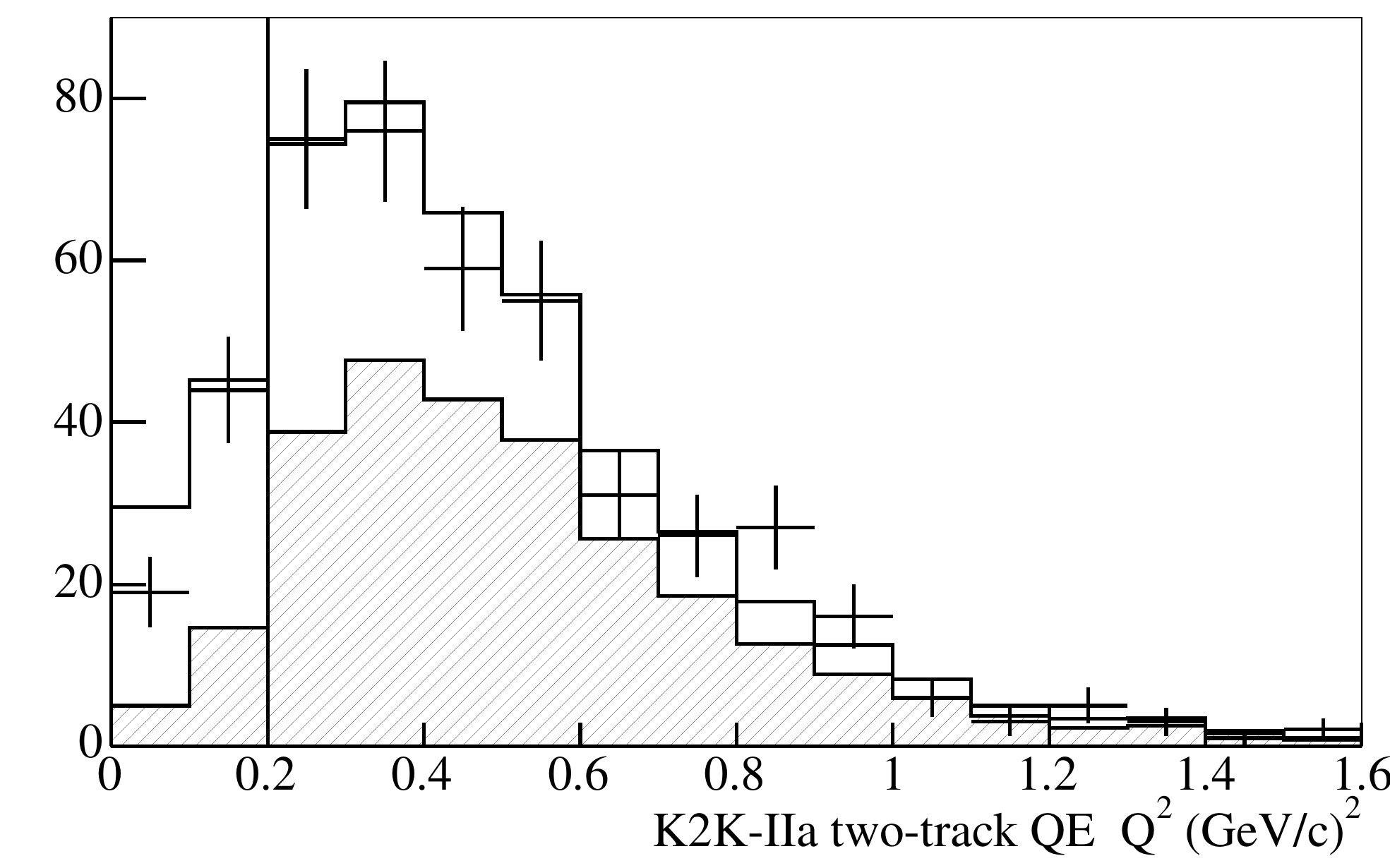}
\includegraphics[width=1.6in,height=1.6in]{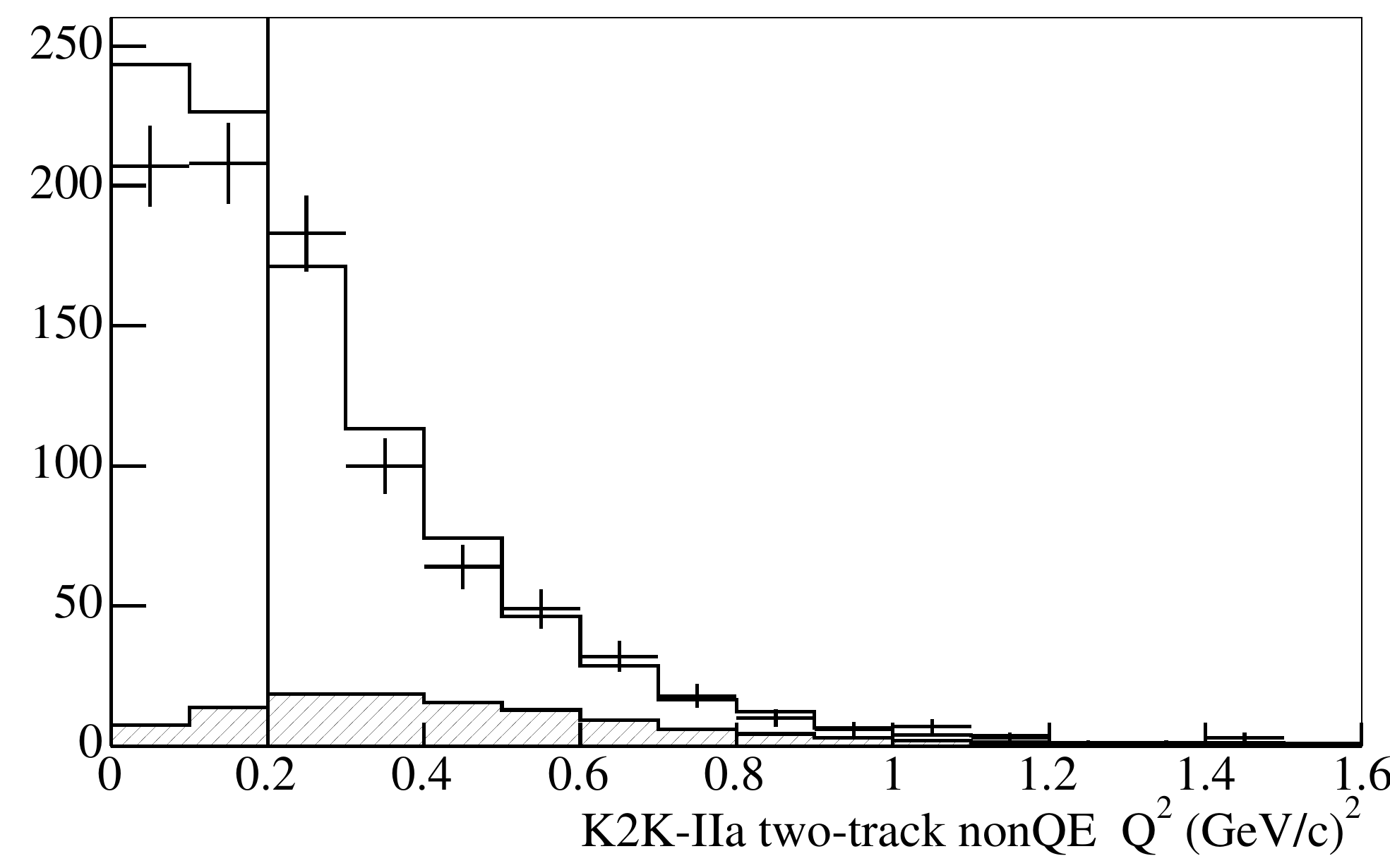}
\caption{\em K2K SciFi \ccqe \, data: reconstructed \q2 distributions for
  K2K-1 data (top) and K2K-IIa data (bottom) for the 1-track, 2-track
  QE enhanced, and 2-track non-QE enhanced samples.  The shaded region
  shows the QE fraction of each sample, estimated from the MC. The
  lowest two data points in each plot are not included in the
  fit.\cite{ccqe_k2k_sf}}
\label{fig:ccqe_k2k_sf}}
\end{figure}

\subsection{K2K SciFi \maqe Analysis}
\label{sec:ccqe_k2k_sf}

The K2K SciFi group published the first \ccqe \, result at these
energies in nearly 20 years~\cite{ccqe_k2k_sf}.  To simulate
neutrino-nuclear scattering, K2K uses the NEUT generator MC
simulation~\cite{neut}; for \ccqe scattering NEUT uses the
Llewellyn-Smith cross section with non-dipole vector form factors and
the Smith-Moniz RFG model.  The SciFi detector is comprised of
multiple aluminium modules each containing a network of scintillating
fibres in water read out by CCD cameras equipped with image
intensifiers. The predominant nuclear target is oxygen.  The fibres
are oriented in the plane transverse to the neutrino beam direction.
Charged particle tracks are detected and their positions and angles
reconstructed by collecting light from the fibres in two views
(``top'' and ``side'').  Charged current neutrino events are tagged by
searching for tracks originating in the SciFi fiducial volume and
penetrating into the MRD.  The analysis requires that muons stop
inside the MRD in order to measure their momenta.

For the \ccqe analysis, events are split into three subsamples based
on the presence and angle of a second track coming from the neutrino
interaction point (defined by the beginning of the muon track).  Only
one and two track events are used in the analysis.  One-track muon
events are grouped together; K2K's Monte Carlo (MC) simulation
indicates that more than 98\% of tagged muon tracks are actually
muons~\cite{ccqe_k2k_sf}.  Two-track events are split into two
subsamples: a QE-enhanced and a non-QE sample.  The QE-enhanced sample
is selected by requiring the second track angle be within 25\deg \, of
the predicted direction based on the observed muon track angle and the
assumption that the event was a \ccqe interaction (\delqp$<$25\deg);
the non-QE sample is the complement and is used to constrain the
backgrounds.  Once the samples are defined, the analysers fit for the
value of \maqe \, by comparing the reconstructed \q2 \, distributions
from data with MC simulation.  The neutrino energy and momentum
transfer are reconstructed using only the observed muon energy and
angle under the assumption that the neutrino event was a \ccqe
interaction.

Figure~\ref{fig:ccqe_k2k_sf} shows the \q2 \, distributions for the
three subsamples described above, both data and MC simulation, broken
into two experimental configurations.  In the K2K-I period, the LG
detector was situated between SciFi and the MRD; in the K2K-II period
the LG was replaced by SciBar.  
The data points below 0.2~\gev2 \, are not used in the fit to avoid
complications from the effects of the nuclear environment.  The
extracted value of \maqe from the fit is 1.20$\pm$0.12~GeV/c$^2$,
which is significantly higher than the average of previous
measurements, 1.015~GeV/c$^2$~\cite{bodek_ma}.  We note that, because
of the dipole form, a high value of \maqe \, does not just affect the
shape of the \q2 \, distribution, it also increases the total rate of
\ccqe events---which does not conflict with the K2K data.

\subsection{K2K SciBar \ccqe Analysis}
\label{sec:ccqe_k2k_sb}

The K2K SciBar \ccqe analysis~\cite{ccqe_k2k_sb} begins in a
similar fashion to the SciFi analysis. SciBar was comprised of 64
layers of fully active scintillating strip planes read out by
multi-anode photomultipliers (MA-PMTs). Each layer contains two planes
of perpendicular strips, with the planes oriented transverse to the
neutrino beam direction. Charged-current neutrino events are first
selected by matching tracks between SciBar and the MRD. Then the data
are split into three samples: one-track, two-track QE enhanced and
two-track non-QE.  Each event in the two-track QE sample has a second
track which satisfies \delqp $<$20\deg , while the non-QE sample
tracks satisfy \delqp $>$20\deg.

For each event, the values of \enu \, and \q2 , assuming a \ccqe
interaction, are reconstructed.  The \q2 \, distributions from the three
samples are fit simultaneously for the value of \maqe .  The best
fit value is \maqe = 1.14$\pm$0.077(fit)$^{+0.078}_{-0.072}$(sys) \gev2 .  This
value for \maqe uses non-dipole vector form factors; the analysers found
that the form of the vector form factors has a significant effect on
the extracted value of \maqe .

\subsection{MiniBooNE \ccqe \, Analysis}
\label{sec:ccqe_mb}

\begin{figure}
\center
{\includegraphics[width=4in]{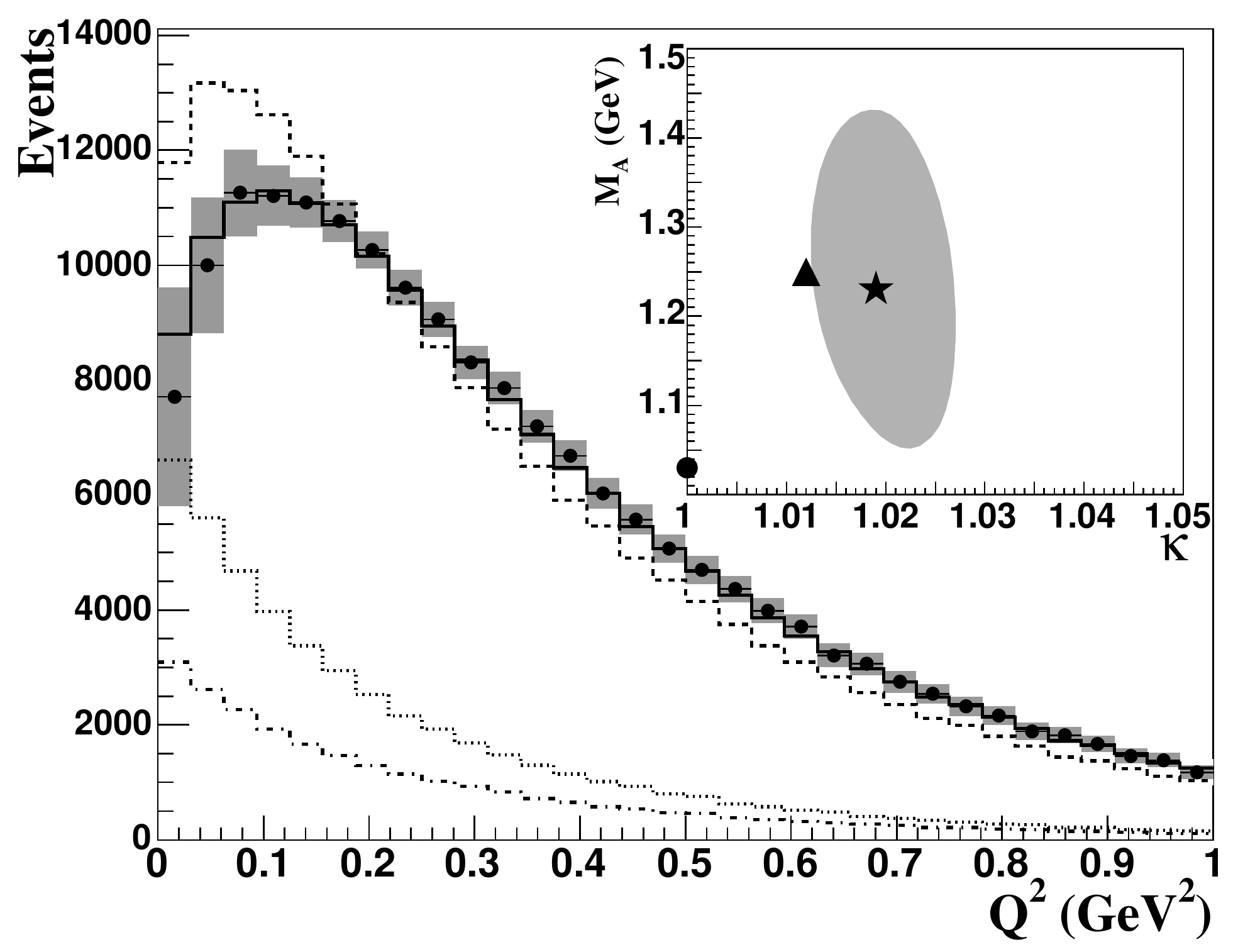}
\caption{\em MiniBooNE \ccqe data: reconstructed \q2 for \numu \ccqe
  events including systematic errors. The simulation, before (dashed)
  and after (solid) the fit, is normalised to data. The dotted
  (dot-dash) curve shows backgrounds that are not \ccqe (not
  ``\ccqe \,-like''). The inset shows the 1$\sigma$ CL contour for the
  best-fit parameters (star), along with the starting values (circle),
  and fit results after varying the background shape
  (triangle).\cite{ccqe_mb}}
\label{fig:ccqe_mb}}
\end{figure}

The MiniBooNE detector is a sphere of mineral oil with 1280 8'' PMTs
at 575~cm radius facing the centre.  The MiniBooNE \ccqe
analysis~\cite{ccqe_mb} begins by selecting clean muon neutrino
events, which are identified by observing the muon's Cherenkov ring
followed by the Cherenkov ring produced by the decay electron.
Requiring the decay electron be located near the end of the
reconstructed muon track yields a high purity \numu \ccqe sample.
A large fraction of background events are charged current single pion
(\ccpip$\!\!$), \numuccpip, interactions in which the final state pion is
not observed.  These \ccqe-like backgrounds can be constrained with a
sample of \ccpip events selected from data by tagging events with two
decay electrons~\cite{cc1pi_nuint05}.

Once the \ccqe \, sample is selected, the analysers examined distributions
of the cosine of the muon angle versus the muon kinetic energy and
found some disagreement in the shapes of the data and MC
distributions.  MiniBooNE uses the \texttt{nuance}~\cite{nuance}
neutrino generator MC, which uses the Llewellyn-Smith cross section
with non-dipole vector form factors and the Smith-Moniz RFG model.  By
plotting the ratio of data over MC, the analysers noted that the shape
disagreement between data and MC followed lines on constant \q2 , not
lines of constant \enu .  This suggests that the source of the
disagreement lay with the cross section model, not the neutrino flux
prediction.

To address the discrepancy, MiniBooNE introduced a new parameter into
the Pauli blocking scheme within the Smith-Moniz RFG model.  The new
parameter, $\kappa$, is a scale factor on the lower bound of the
nucleon sea and controls the size of the nucleon phase space relevant
to Pauli blocking.  Then MiniBooNE performed a fit to the
reconstructed \q2 \,distribution (as in the K2K analyses, both \enu \,
and \q2 \, are reconstructed assuming a \ccqe \, interaction) to
extract the value of \maqe and $\kappa$.  The best fit values are
\maqe = 1.23$\pm$0.20~GeV/c$^2$ and $\kappa$ = 1.019$\pm$0.011.

The effect of the fit is given in figure~\ref{fig:ccqe_mb}.  The
pre-fit MC curve lies above the data at low \q2 $\!$, where Pauli
blocking occurs, and below the data at high \q2 $\!$.  After the fit,
the MC agrees the data across the whole range of \q2 $\!$.  The high
value of \maqe causes a harder \q2 \, spectrum, which improves
agreement at high \q2 $\!$, and the increased Pauli blocking caused by
the high value of $\kappa$ suppresses production at low \q2 $\!$.  As
mentioned in section~\ref{sec:ccqe_k2k_sf} a high value of \maqe also
increases the total event rate.  Nevertheless the ratio of MiniBooNE's
observed \ccqe event rate to predicted (using the best fit parameters)
is 1.21$\pm$0.24~\cite{ccqe_mb}.

\subsection{SciBooNE \ccqe \, Analysis}

\begin{figure}
\center
{\includegraphics*[width=4.9in,height=3.5in,viewport=5 110 780 470]{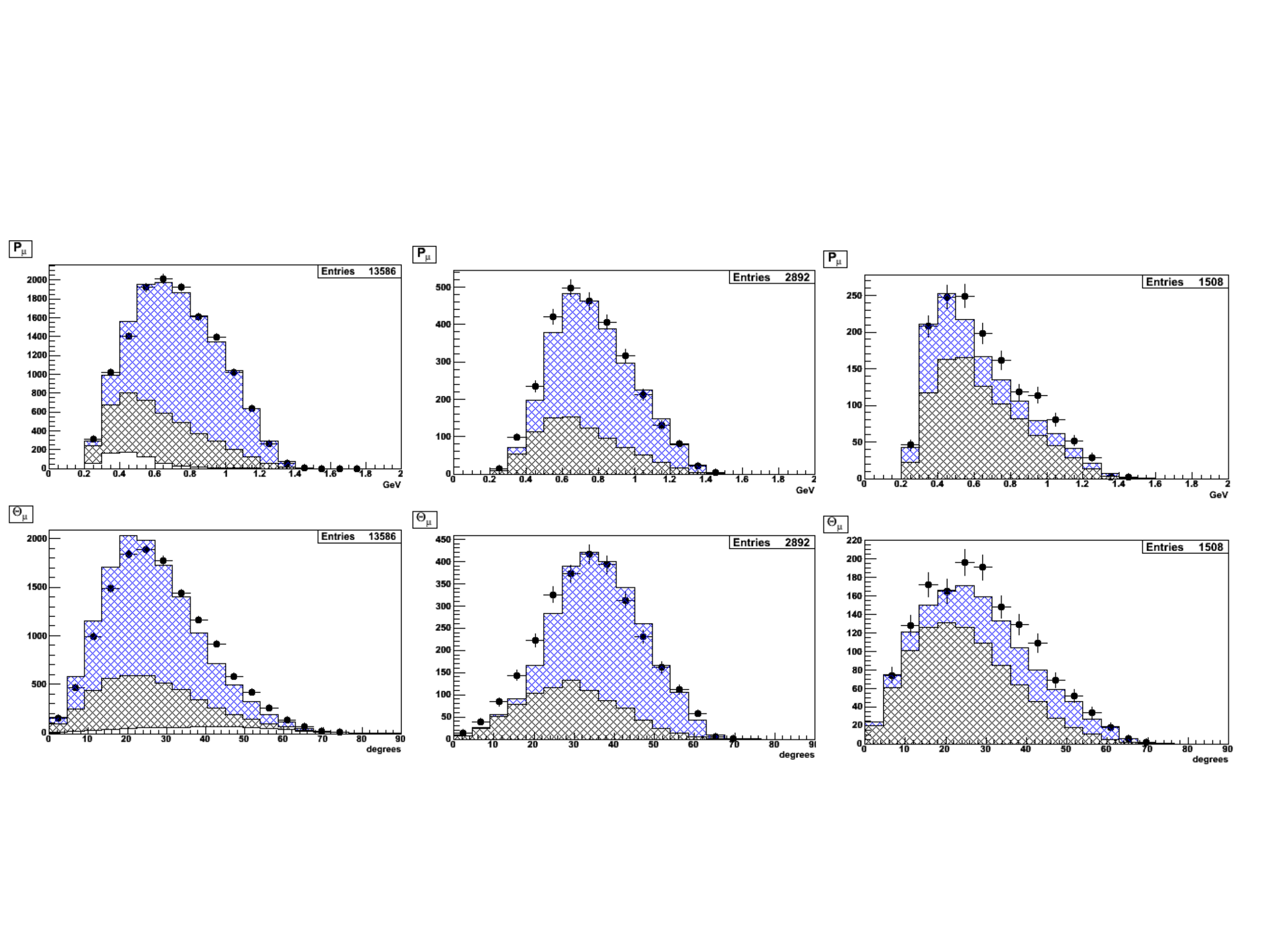}
\caption{\em SciBooNE \ccqe data: muon momentum (top) and angles
  (bottom) for SciBar-MRD matched sample.  The left panels show
  one-track events, the centre panels show two track \mup events and
  the right panels show two-track \mupi events.  The data (points)
  include statistical uncertainties only.  The MC (histogram) is split
  into three components: \ccqe (blue), non-QE (black) and events
  originating outside the SciBar fiducial volume (white).  The MC was
  generated with \maqe=1.21~GeV/c$^2$ and is normalised to the
  MRD-matched data.\cite{ccqe_sb}}
\label{fig:ccqe_sb_mrd}}
\end{figure}

SciBooNE is developing two distinct \ccqe data sets, one with tracks
matched between SciBar and the MRD and the other with tracks contained
within SciBar.  To simulate neutrino-nuclear scattering, SciBooNE uses
the NEUT generator Monte Carlo (MC) simulation~\cite{neut}.

\begin{figure}
\center
{\includegraphics*[width=4.9in,viewport=40 40 690 460]{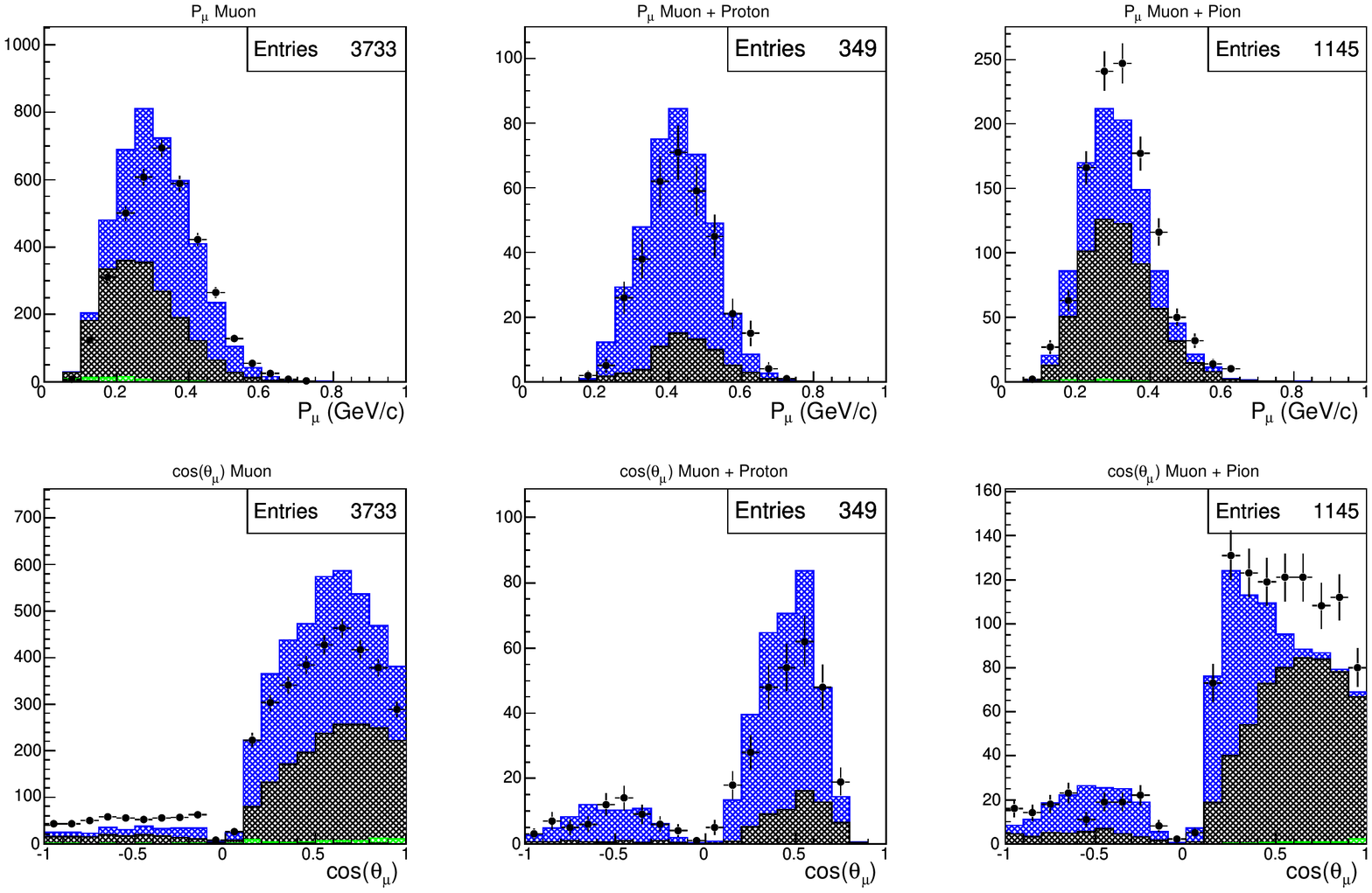}
\caption{\em SciBooNE \ccqe data: muon momentum (top) and angles
  (bottom) for SciBar contained sample.  The left panels show
  one-track events, the centre panels show two track \mup events and
  the right panels show two-track \mupi events.  The data (points)
  include statistical uncertainties only.  The MC (histogram) is split
  into three components: \ccqe (blue), non-QE (black) and events
  originating outside the SciBar fiducial volume (green).  The MC was
  generated with \maqe=1.21~GeV/c$^2$ and is normalised to the
  SciBar-MRD matched data.\cite{ccqe_sb}}
\label{fig:ccqe_sb_sbcon}}
\end{figure}

Charged-current neutrino candidates in the MRD sample are selected by
matching tracks originating in the fiducial volume of SciBar and
penetrating into the MRD; the muons are tagged by their penetration
into the MRD.  The upstream end of the muon track defines the neutrino
interaction vertex. The analysers separate events based on the number
of tracks coming out of the neutrino interaction vertex.  One track
events have no tracks other than the muon candidate.  It was found
that there was significant disagreement between data and MC in the
\delqp \, distributions, so that parameter was not used to separate
signal QE from background non-QE events.  Instead, two track events
are separated into \mup and \mupi samples using particle
identification based on the energy deposited along the second track.
The one-track and \mup samples are predominantly \ccqe events, and the
\mupi sample is predominantly \ccpip
events. Figure~\ref{fig:ccqe_sb_mrd} shows the data and MC
distributions for the SciBar-MRD matched sample.

In the SciBar-contained sample, the muons from charged-current
neutrino candidates are tagged with particle identification based on
energy deposit along the track and by searching for their decay
electrons using SciBar's multi-hit TDCs~\cite{sb_nuint07}.  Events are
further classified based on the number and type of tracks coming from
the neutrino vertex. Removing the MRD track-matching cut allows the
reconstruction of backwards-going tracks, thus expanding the \q2 \,
range open to the analysis. The neutrino vertex is defined using the
timing of hits within the muon track, and the location of the tagged
decay electron. The SciBar contained sample is split into one-track
muon events, two track \mup and two track \mupi
events. Figure~\ref{fig:ccqe_sb_sbcon} shows the data and MC
distributions for the SciBar-contained sample.  The analyses are
ongoing.x


\section{Charged Current Single Pion Production}
\label{sec:cc1pi}

The charged-current single pion (\ccpip $\!$) production processes,
\numuccpip, are the second most copious near 1~GeV neutrino energy.
They offer rich phenomenology compared to the quasi-elastic process
but because there is only one additional final state particle they are
simple to tag and reconstruct experimentally.  In oscillation
experiments they form the primary background channel in \numu \,
disappearance searches; the final state pion can be absorbed into the
nuclear medium and hence escape observation in the neutrino detector,
forming an irreducible background.  This phenomenon makes a precise
understanding of \ccpip scattering a high priority for the next
generation of oscillation experiments~\cite{itow-debbie}.

Single pion production on nuclei is often broken into two broad
phenomenological categories, coherent and incoherent scattering.

The creation of resonances via interaction of the neutrino with a
single nucleon dominates pion production near 1~GeV, and is broadly
referred to as incoherent scattering. The most commonly used model for
predicting the \ccpip cross section, and kinematics of final state
particles, is the Rein and Sehgal (RS) model~\cite{rein-sehgal}.  The model
is attractive because it describes all neutrino and antineutrino pion
production processes in one uniform framework.  (The RS model is used
to predict neutral current single pion production processes as well.)
The model is based on the Feynman, Ravndal and Kislinger~\cite{fkr}
relativistic quark oscillator approach and includes the excitations of
18 resonances below hadronic invariant mass 2 GeV/c$^2$.  The model is
parameterised by form factors that are usually assumed to have a
dipole form dependent on mass parameters similar to the dipole forms
assumed for the axial form factor in \ccqe scattering.

Coherent pion production is the interaction of the neutrino with the
nucleus as a whole, or in other words with all the nucleons
coherently, to produce a pion, $\nu_{\mu}A\rightarrow\mu^-A\pi^+$.
The scattering process leaves the nucleus in its ground state, so is
an inherently low momentum-transfer process.  Again, the most commonly
used model to predict the coherent production of pions is the model of
Rein and Sehgal~\cite{rein-sehgal-coh}.

\subsection{K2K \ccpip Analysis}
\label{sec:cc1pi_k2k}

\begin{figure}
\center
{
\includegraphics[width=2.35in]{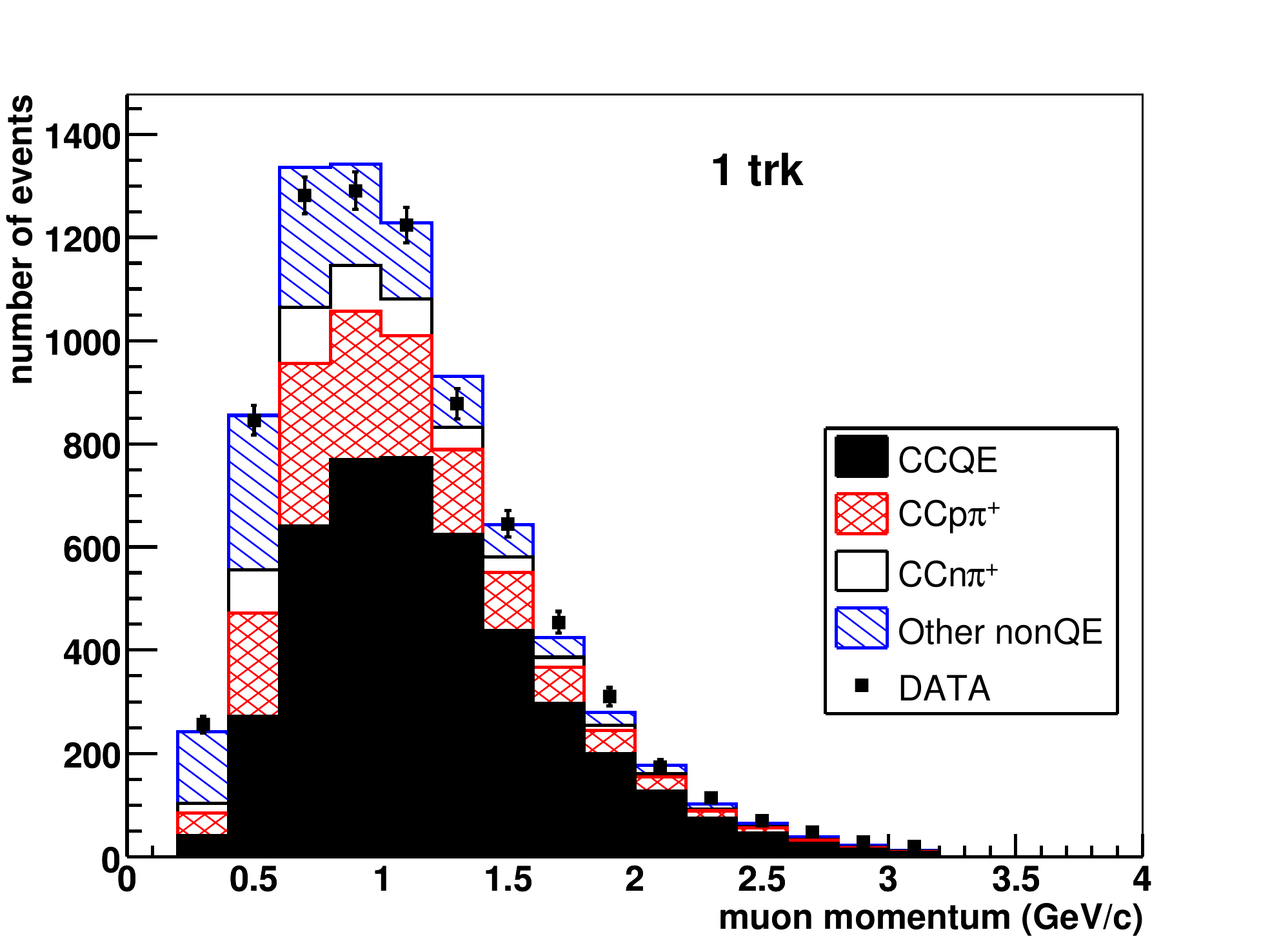}
\includegraphics[width=2.35in]{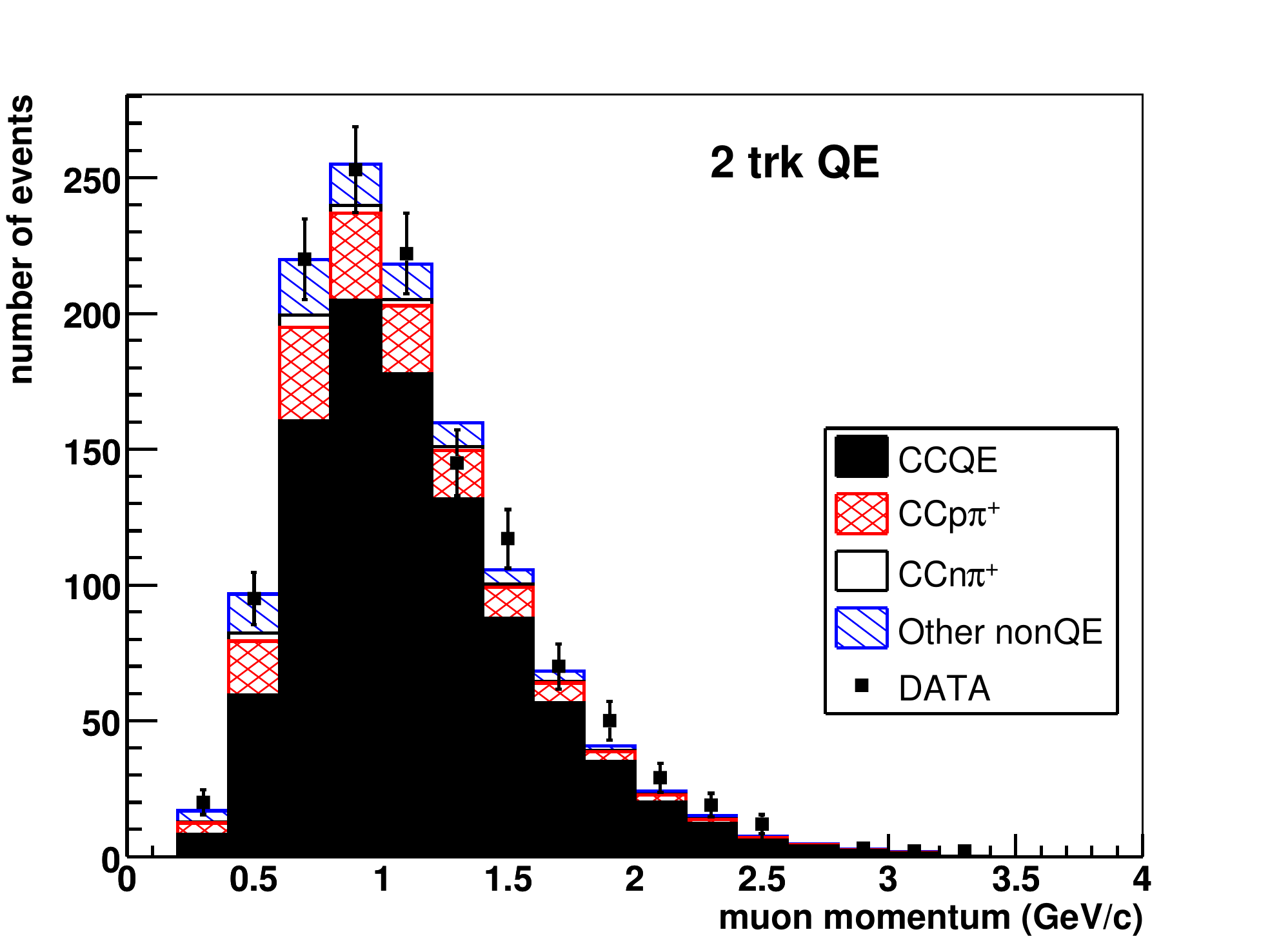}
\includegraphics[width=2.35in]{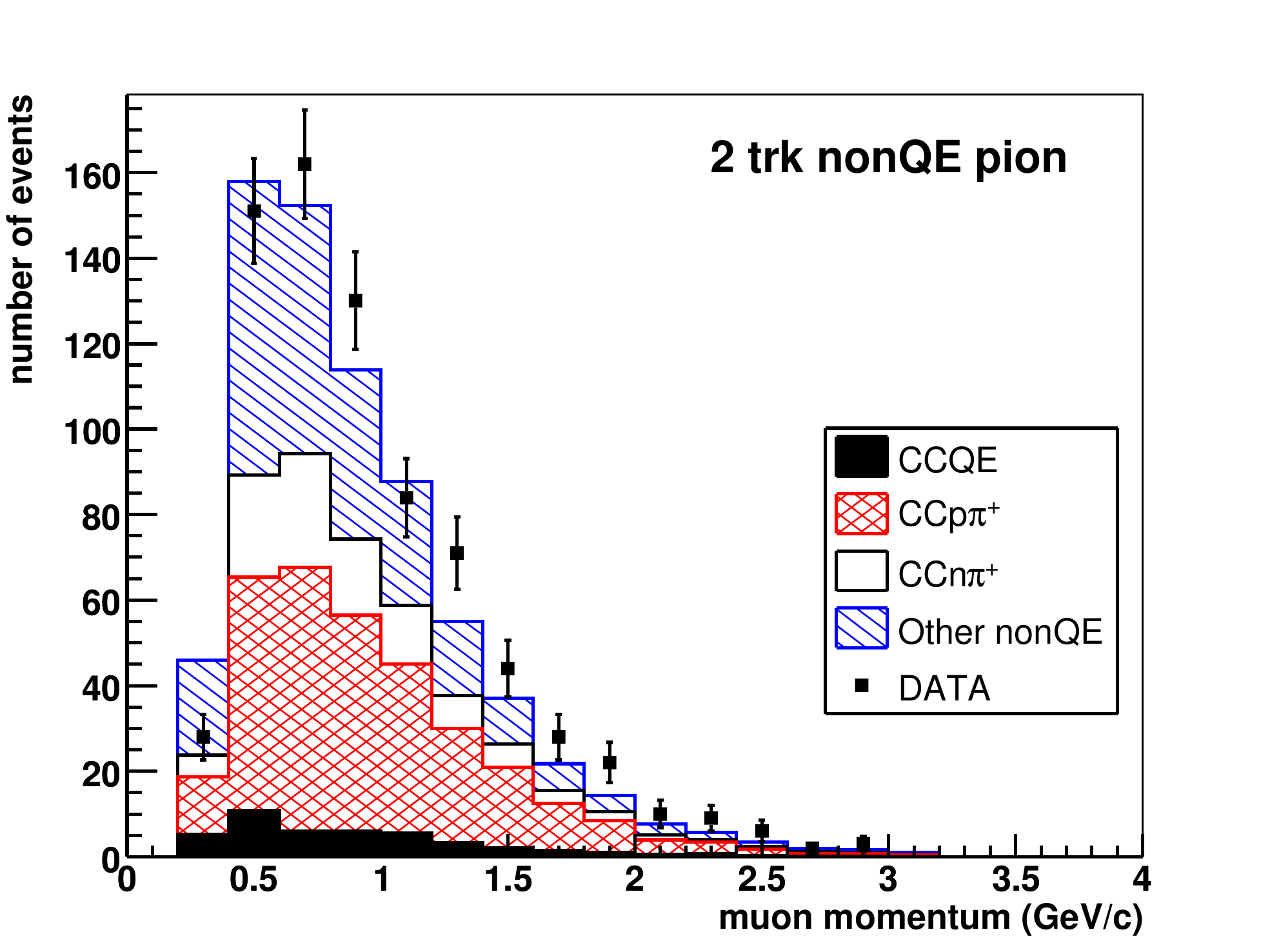}
\includegraphics[width=2.35in]{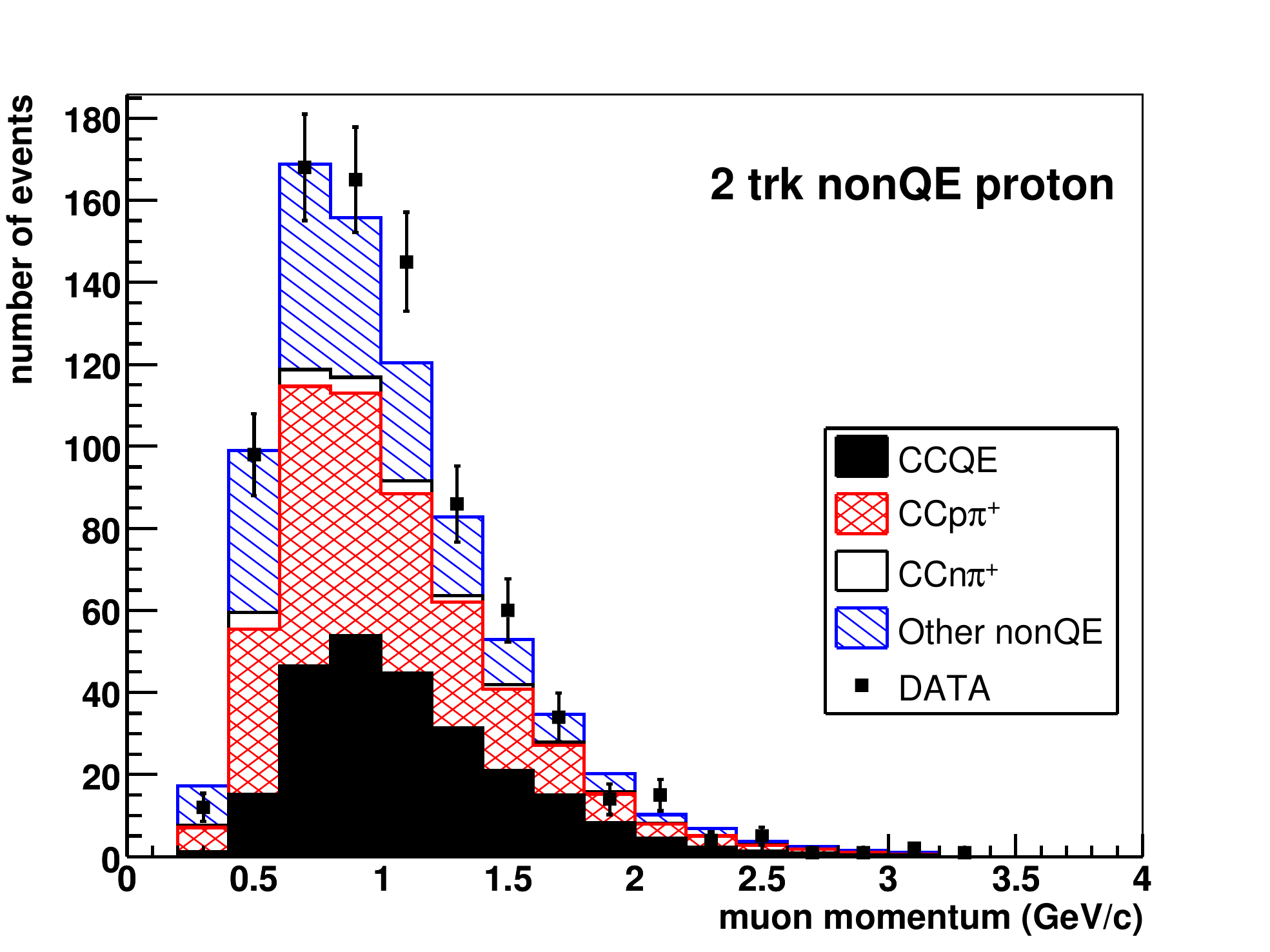}
\caption{\em K2K SciBar \ccpip data; shown are the muon momentum
  distributions for each of the four data samples used in the
  analysis.  In the top left are the one-track events, in the top
  right are the two track QE events, in the bottom left are the two
  track non-QE \mup events and in the bottom right are the two
  track non-QE \mupi events.  In each panel, the data (black points)
  are shown with statistical uncertainty, along with \ccqe (black),
  \ccpip (red and white) and other non-QE (blue) contributions from
  the MC.\cite{cc1pi_k2k}}
\label{fig:cc1pi_k2k}}
\end{figure}

K2K's \ccpip analysis goal was an extraction of the energy dependent
ratio \ccpipratio~\cite{cc1pi_k2k}. Again, K2K uses the NEUT MC
generator; NEUT models \ccpip interactions using the RS model.  The
analysis begins with CC neutrino candidates found by matching tracks
between SciBar's fiducial volume and the MRD.  In such events, the
muon is required to stop in the MRD to afford a good muon momentum
measurement.  The analysis uses one-track and two-track events
only. Two track events are split into QE and non-QE samples with a
\delqp \, cut at 20\deg.  The non-QE sample is split again into a \mup
and a \mupi sample using particle identification based on energy
deposition along the track.  The resultant four samples are used to
fit for the relative fractions of \ccqe, \ccpip and other non-QE
events, and thereby extract the cross section ratio.  The muon
momentum distributions are shown in figure~\ref{fig:cc1pi_k2k}.

The analysis proceeds by performing a maximum likelihood fit of the
data and MC in bins of $p_{\mu}$ and $\theta_{\mu}$ over the four
samples simultaneously.  The MC is split into four true neutrino
energy bins to extract information on the energy dependence of the
\ccpipratio \, ratio.  The result of the analysis is shown in
figure~\ref{fig:cc1pi_compare}~(right) compared with the MiniBooNE and
Argonne National Lab~\cite{cc1pi_anl} \ccpipratio results.

\subsection{MiniBooNE \ccpip Analysis}
\label{sec:cc1pi_mb}

\begin{figure}
\center
{
\includegraphics[width=2.4in]{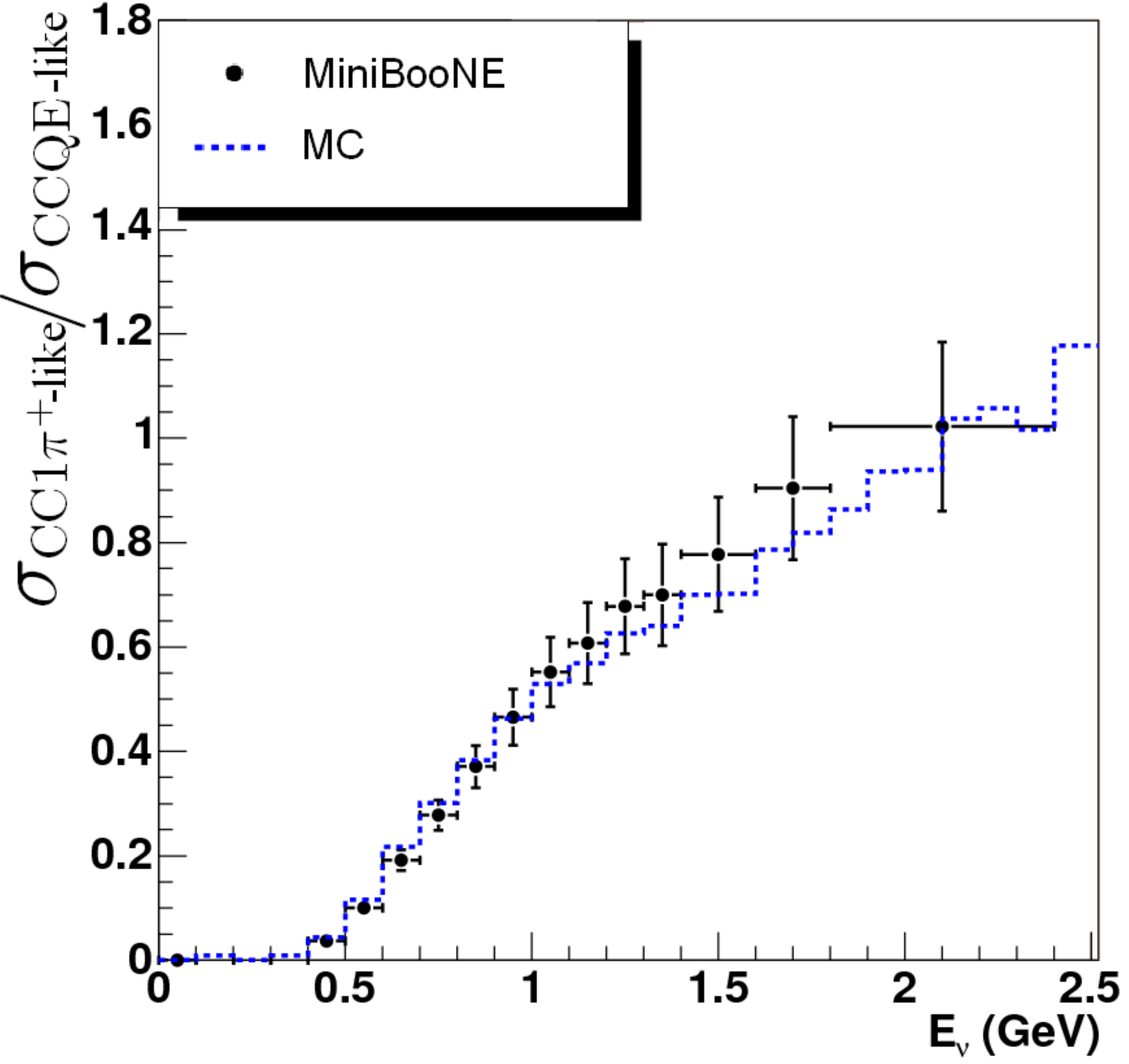}
\includegraphics[width=2.4in]{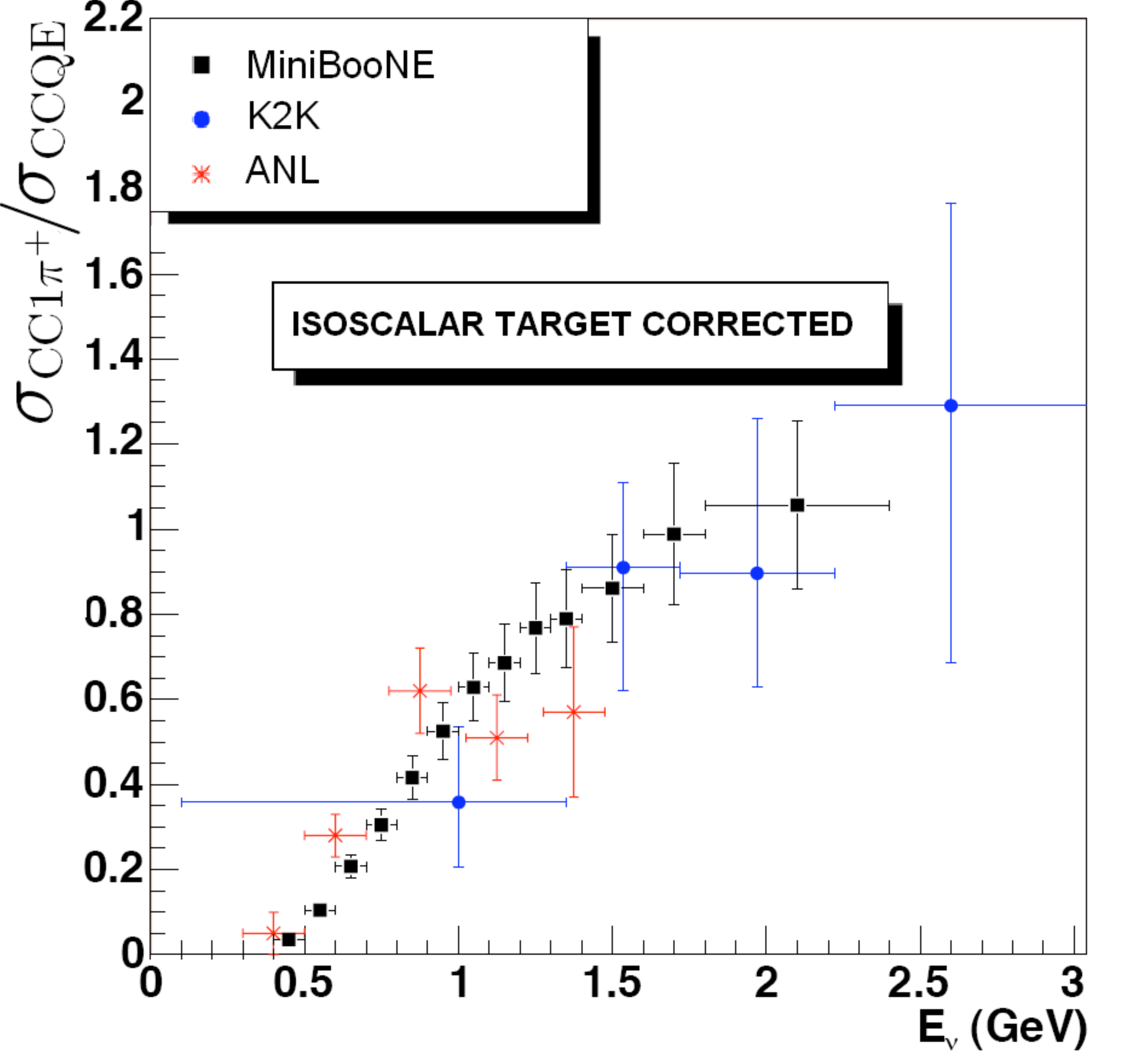}
\caption{\em MiniBooNE \ccpip$\!\!$/\ccqe \, ratio measurement (left) and
  comparison of MiniBooNE data with results from K2K and
  ANL.\cite{cc1pi_mb_prl}}
\label{fig:cc1pi_compare}}
\end{figure}

As mentioned in section~\ref{sec:ccqe_mb}, MiniBooNE selects a high
purity sample of \ccpip events by tagging neutrino events with two
decay electrons.
Using this simple cut MiniBooNE extracts an high statistics and purity
\ccpip data set. The predominant source of Cherenkov light in the
MiniBooNE \ccpip events is the $\mu^-$, so the events are fitted with
the standard single ring reconstruction algorithm to find $E_{\mu}$ and
$\theta_{\mu}$.  Those values are used to calculate \enu \, and \q2 \,
assuming \ccqe(2 body) kinematics but assuming that the recoiling
particle has the mass of the $\Delta$(1232) resonance instead of the
mass of a nucleon.  The energy dependent cross section ratio is
calculated directly by using the MiniBooNE \ccqe data for the
denominator.

MiniBooNE uses the \texttt{nuance} MC generator which models \ccpip
production using the RS model.

\begin{figure}
\center
{
\includegraphics[width=4.in]{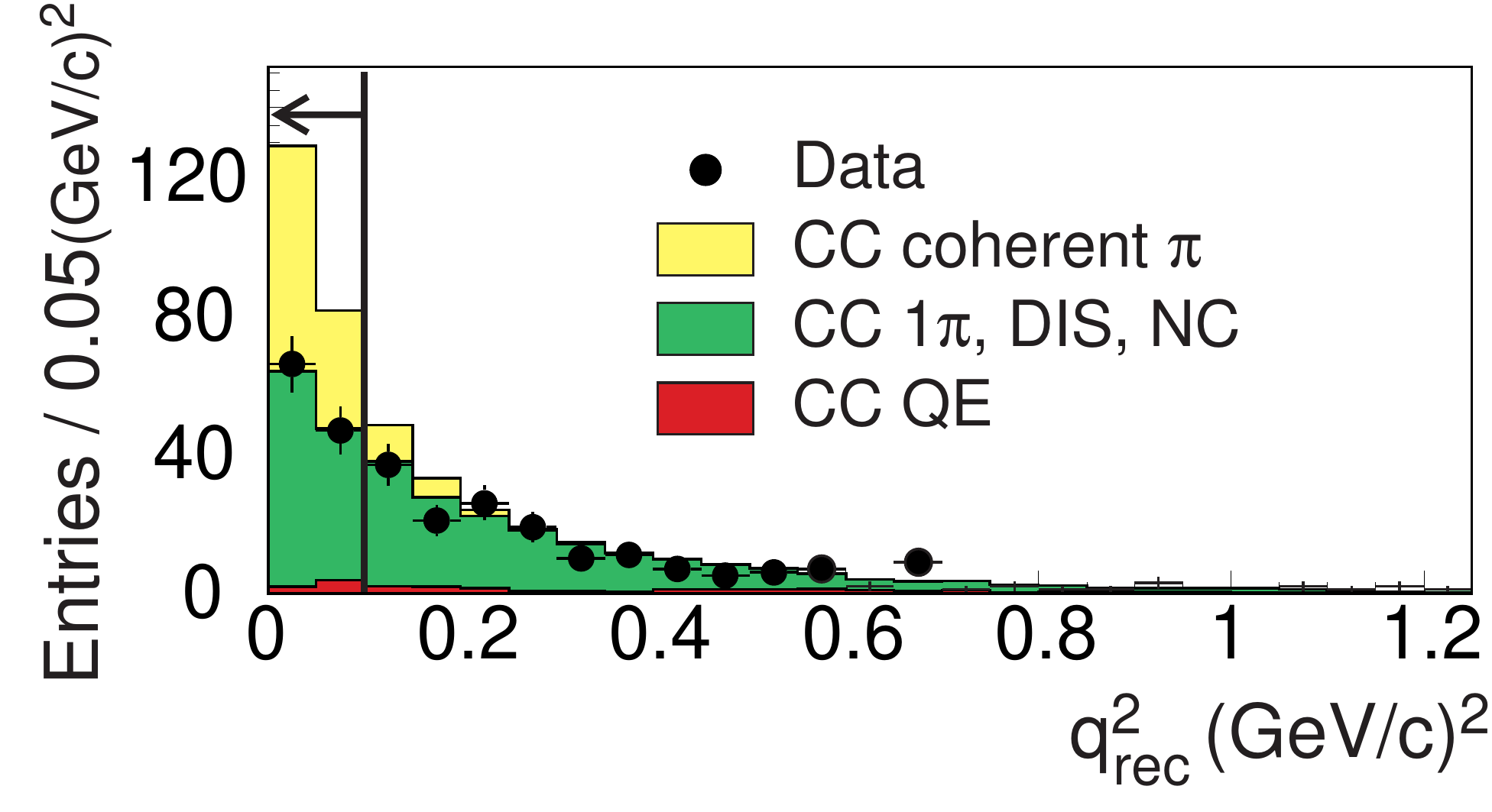}
\includegraphics[width=4.in]{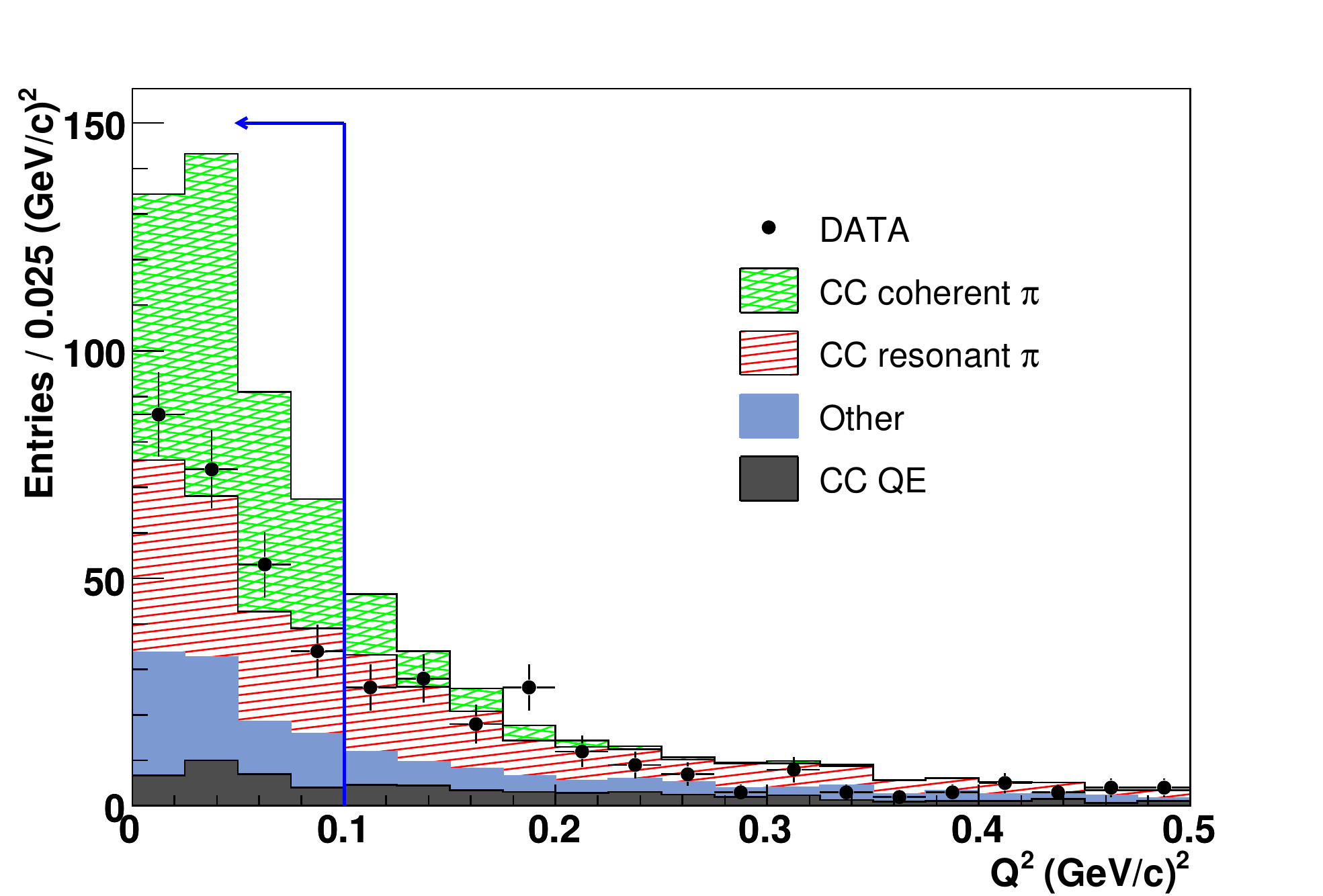}
\caption{\em K2K and SciBooNE neutrino coherent pion search data.  Each
  plot shows reconstructed \q2 \, distributions for the coherent pion
  enhanced samples, which are \mupi \, non-QE events with low vertex
  activity.  The top panel shows K2K SciBar data and best fit
  MC~\cite{cccohpi_k2k}, and the bottom panel shows SciBooNE data with
  best fit MC~\cite{cccohpi_sb}.}
\label{fig:cccohpi_q2}}
\end{figure}

MiniBooNE calculates the cross section ratio in two
ways~\cite{cc1pi_mb_prl}.  The primary measurement is
\ccpiplikeratio$\!$, a so-called {\em effective} cross section
ratio. \ccpip$\!\!$-like is defined as an event in which exactly one
$\mu^-$ and one $\pi^+$ exit the struck nucleus, and \ccqe$\!\!$-like
is defined as an event with exactly one $\mu^-$ and zero pions.  The
{\em effective} \ccpip$\!\!$/\ccqe ratio is shown in
figure~\ref{fig:cc1pi_compare}~(left) as a function of neutrino energy.
The secondary MiniBooNE \ccpip measurement is corrected for final
state interactions (FSI)---mainly hadronic interactions within the
nucleus.  The FSI-corrected ratio is shown in
figure~\ref{fig:cc1pi_compare}~(right) compared with the results from
K2K and ANL.  The {\em effective} ratio does not attempt to make any
corrections for nuclear effects and is thus less model-dependent
than the FSI-corrected cross section ratio.

\subsection{CC Coherent Pion Production}
\label{sec:cc1pi_coh}

Noticing that there was disagreement between data and MC in the
inelastic CC data samples from their near detector FGDs at low \q2 ,
and knowing that coherent pion production is an inherently low \q2 \,
process, the K2K collaboration was inspired to explore CC coherent
pion production with the SciBar detector.

Starting with tracks matched between SciBar and the MRD, the analysers
use two-track events to select a sample of CC coherent pions. The
two-track sample is split into QE and non-QE samples using a cut on
\delqp \, at 25\deg. The non-QE sample is further refined by using the
energy deposited along the track to distinguish \mup from \mupi
events.  The analysers tune the MC simulation using the two-track
non-QE \mupi data sample, which is enriched with signal events, and
the complementary data samples: one-track, two-track QE, and two-track
non-QE \mup, which are all background enriched samples.  The tuning is
done by fitting the MC reconstructed \q2 \, distributions to the data;
in all samples \q2 \, is reconstructed under the assumption of \ccqe
\, kinematics.  Once the MC has been tuned, the final event selection
cuts are made on the non-QE \mupi sample.  Events are required to have
forward-going pions. To remove events in which a third particle is
emitted but cannot be reconstructed as a track, events are required to
have less than 7~MeV {\em vertex activity:} energy deposited in the
scintillator strip which contains the neutrino vertex. The
reconstructed \q2 \, distributions (data and MC) are shown in
figure~\ref{fig:cccohpi_q2}~(top). The CC coherent signal sample uses
only events with reconstructed \q2 \, below 0.1~\gev2 ; as seen in
figure~\ref{fig:cccohpi_q2}, the K2K data are consistent with the
background prediction.  The analysers therefore set an upper limit on
the CC coherent pion to CC inclusive cross section ratio, $\sigma(CC
coh \pi)/\sigma(CC inc)<0.60\times10^{-2}$ at 90\% CL.

The SciBooNE CC coherent pion search proceeds along similar lines to
the K2K search.  Four data samples are selected and used to tune the
MC simulation before making the final CC coherent pion event
selection~\cite{cccohpi_sb}.  In the SciBooNE analysis, the four
samples used to tune the MC are the one-track, two-track \mup,
two-track \mupi with high vertex activity and two-track \mupi with low
vertex activity.  Again, the MC tuning is done with reconstructed \q2
\, distributions.  Once the MC is tuned, the final event sample is
made by rejecting \ccqe events with a \delqp \, cut at 20\deg \, and
by requiring forward-going pions. The reconstructed \q2 \,
distributions (data and MC) are shown in
figure~\ref{fig:cccohpi_q2}(bottom). The CC coherent signal sample
uses only events with reconstructed \q2 \, below 0.1~\gev2 ; as seen in
figure~\ref{fig:cccohpi_q2}, the SciBooNE data are consistent with the
background prediction.  The analysers therefore set 90\% CL upper
limits on the CC coherent pion to CC inclusive cross section ratio,
$\sigma(CC coh \pi)/\sigma(CC inc)<0.67\times10^{-2}$ at 1.1~GeV
neutrino energy and $\sigma(CC coh \pi)/\sigma(CC
inc)<1.36\times10^{-2}$ at 2.2~GeV.


\section{Neutral Current Single Pion Production}
\label{sec:ncpi0}

Neutral current $\pi^0$ production (\ncpi), \numuncpi, is an important
process for experiments searching for $\nu_{\mu}\!\rightarrow\!\nu_e$
oscillations because it accounts for the largest single
misidentification background.  The $\nu_e$ background events arise if
one of the photons from the $\pi^0$ decay is not observed in the
neutrino detector or if the photon tracks overlap closely in the lab
frame; a single $\gamma$ (or overlapping $\gamma$s) creates a shower
that is very nearly indistinguishable from a true electron shower in
an open volume Cherenkov detector.  On the other hand, an open volume
Cherenkov detector is quite good for detecting the majority of
$\pi^0$s produced in its fiducial volume because of the number of
radiation lengths of material presented to the decay photons.  As we
shall see, K2K and MiniBooNE collected remarkably large \ncpi data
sets.

\ncpi production cross sections are typically modeled by experiments
using the RS model.  Because the neutrino carries away an unknown
amount of energy in an NC interaction, the only observables in the
detector are the pion and nucleon kinematic variables (although the
nucleon is usually not observed).

\subsection{K2K \ncpi Analysis}
\label{sec:ncpi0_k2k}

\begin{figure}
\center
{
\includegraphics[width=2.3in]{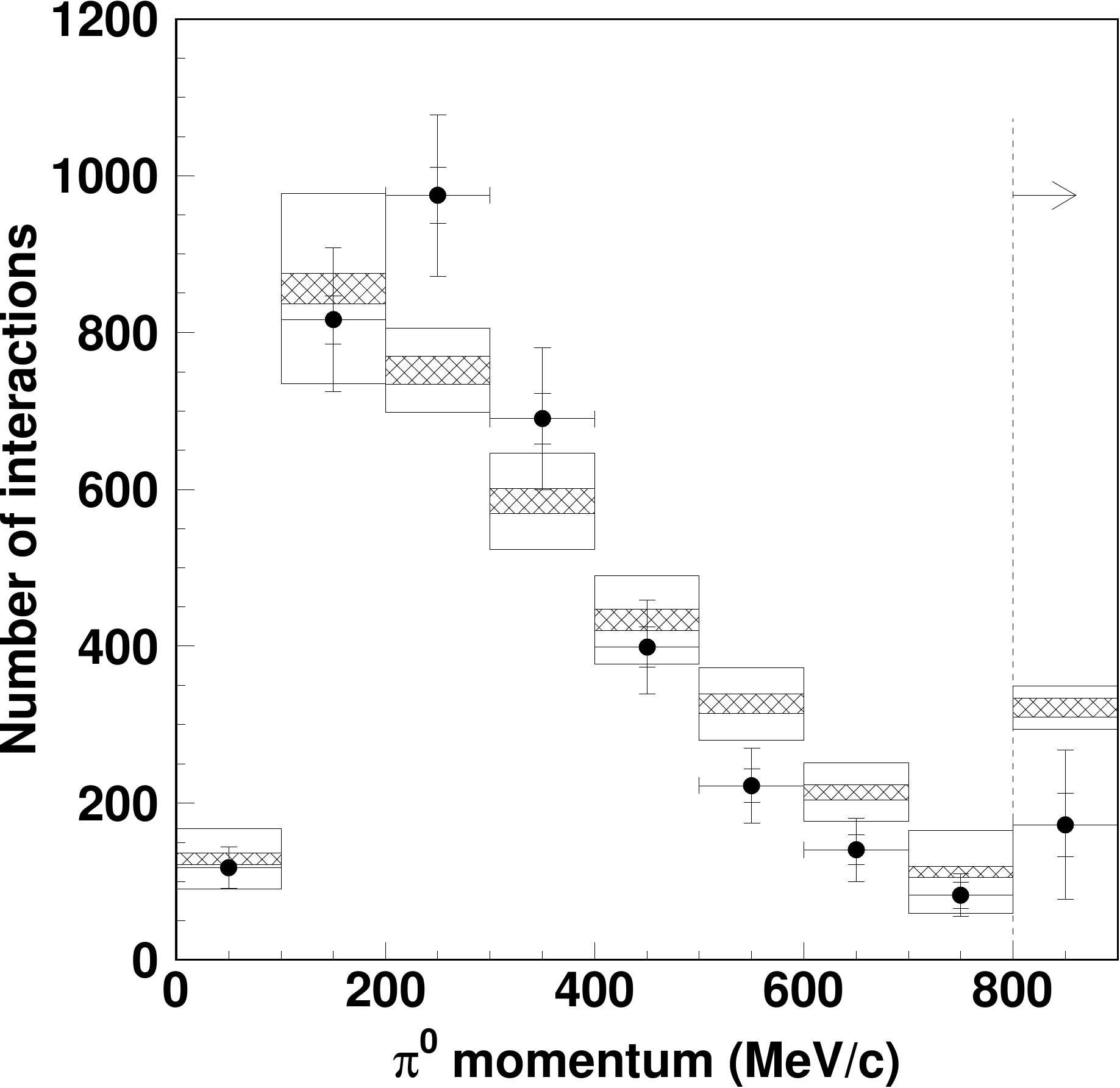}
\includegraphics*[width=2.4in,height=2.3in,viewport=260 5 530 250]{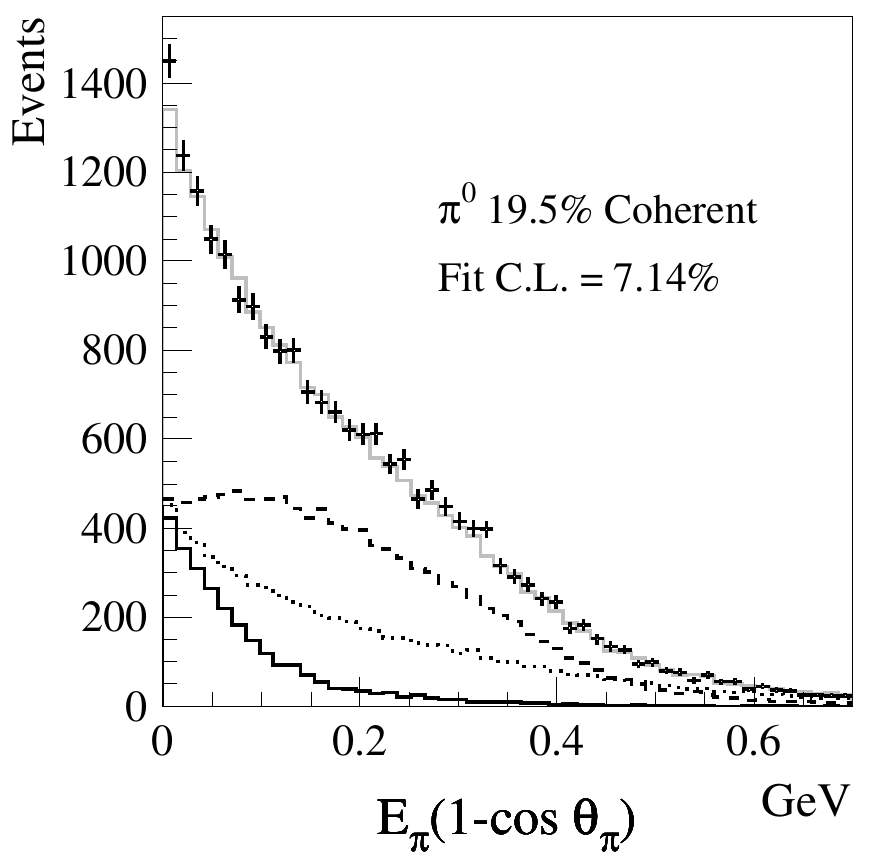}
\caption{\em K2K and MiniBooNE \ncpi data.  Left: momentum
  distribution of K2K \ncpi data (black dots).  The inner and outer
  error bars attached to data points show statistical errors and total
  errors including systematic errors, respectively.  The MC simulation
  is also shown as a box histogram; the inner boxes represent MC
  statistical errors and the outer boxes represent the cross section
  model uncertainty.  Right: energy weighted angular distribution of
  MiniBooNE data (crosses) and the histograms showing the coherent
  (solid), resonant (dashed) and background (dotted) contributions as
  predicted by MC after fitting to the
  data~\cite{ncpi0_k2k,ncpi0_mb}.}
\label{fig:ncpi0}}
\end{figure}

K2K used the 1kT water Cherenkov detector to constrain \ncpi
production for their $\nu_e$ appearance search. \ncpi events are
tagged by requiring: no decay electron (which would be observed as a
delayed signal); a fully contained event (whose signature is no single
PMT with a very high charge hit); two electron-like (showering) rings;
and the reconstructed invariant mass, $M_{\gamma\gamma}$, in the range
85-215~MeV/c$^2$.  The analysers measure the rate and momentum
spectrum of the $\pi^0$s, as well as the \ncpi to CC inclusive cross
section ratio.  The pion momentum spectrum is shown in
figure~\ref{fig:ncpi0}~(left); the measured spectrum is used to
tune the MC prediction of $\pi^0$ production, which significantly
improves the prediction of $\nu_e$ background events for the
appearance oscillation search.  The measured cross section ratio is
$\sigma_{NC1\pi0}/\sigma_{CC-inc}$ =
0.064$\pm$0.001(stat)$\pm$0.007(syst), which agrees well the MC
prediction of 0.065~\cite{ncpi0_k2k}.

\begin{figure}
\center
{\includegraphics[width=4.in,height=3.0in]{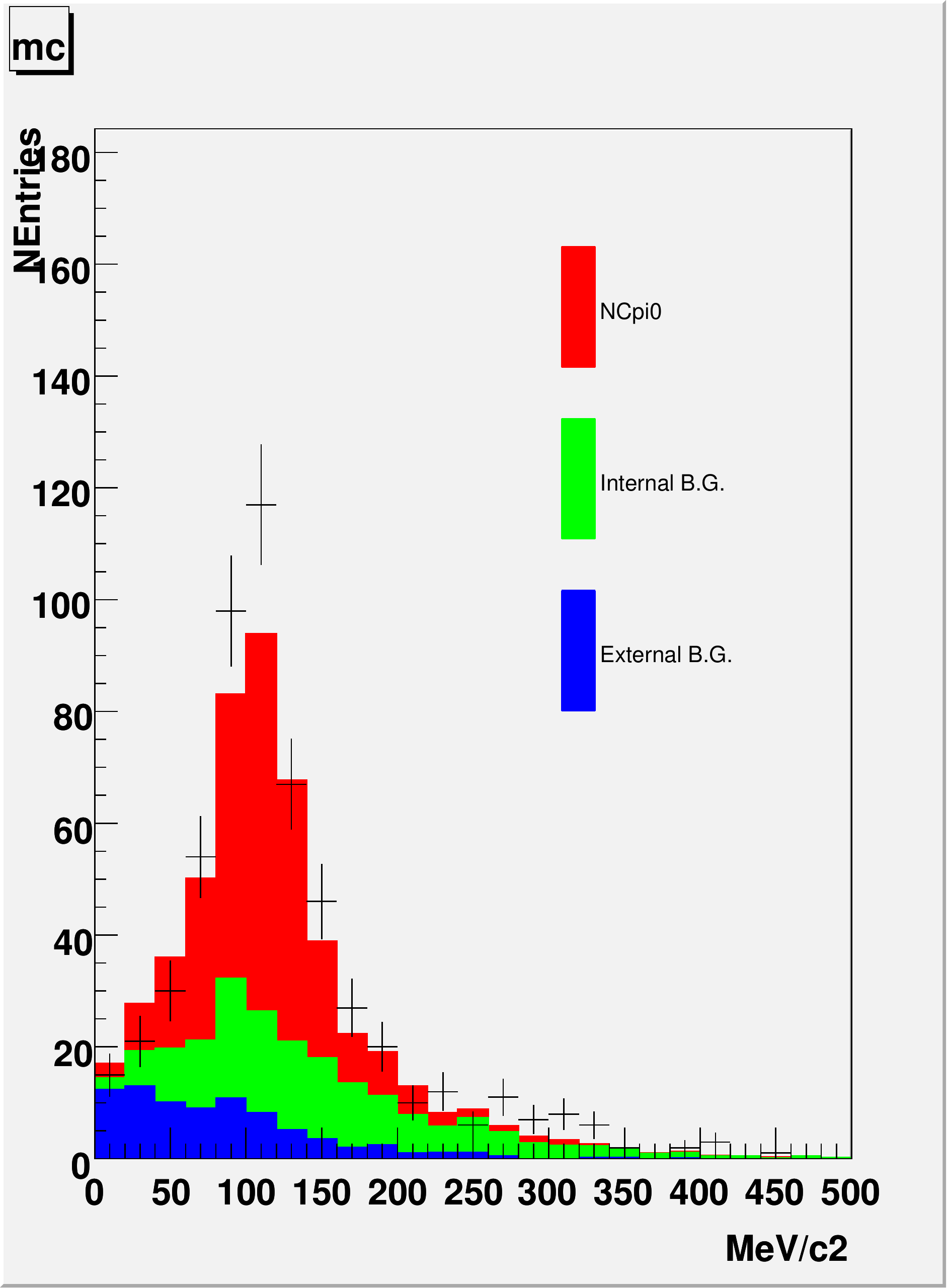}
\caption{\em SciBooNE's econstructed invariant mass for \numu \ncpi
  candidates contained within SciBar~\cite{ncpi0_sb}.  Data (crosses)
  are shown with statistical errors as well as signal \ncpi (red),
  internal background (green) and external background (blue)
  contributions predicted by the MC simulation.}
\label{fig:ncpi0_sb}}
\end{figure}

\subsection{MiniBooNE \ncpi Analysis}
\label{sec:ncpi0_mb}

MiniBooNE was designed to search for $\nu_{\mu}\rightarrow\nu_e$
oscillations, so constraining the \ncpi background measurement is a
crucial part of the experiment's goals~\cite{ncpi0_mb}.  MiniBooNE
analysers select fully contained neutrino candidates with no decay
electron.  The remaining events are reconstructed according to three
separate hypotheses: single muon track, single electron shower and
$\pi^0$ shower.  Likelihood ratios of the three hypotheses are used as
particle identification to select $\pi^0$ candidate events.  The
\texttt{nuance} MC simulation used by MiniBooNE models \ncpi
production with the RS model, and indicates that the signal to
background ratio after the PID cuts is $\sim$30.  Next, the analysers
require 80~MeV$<$M$_{\gamma\gamma}\!\!<$200~MeV and perform a momentum
unsmearing.  The extracted pion momentum spectrum is used to tune the
MC prediction of the $\nu_e$ backgrounds for the oscillation search.
MiniBooNE has also measured the coherent fraction via a template fit
using MC predicted shapes for the coherent, incoherent (resonant) and
other processes in the variable $E_{\pi}(1-\cos\theta_{\pi})$.  The
results of the fit are shown in figure~\ref{fig:ncpi0}~(right); the
extracted coherent fraction observed by MiniBooNE is
(19.5$\pm$1.1(stat)$\pm$2.5(syst))\%~\cite{ncpi0_mb}.

\subsection{SciBooNE \ncpi Analysis}

SciBooNE begins the \ncpi event selection by requiring two tracks
originating in the SciBar fiducial volume disconnected from each other
(due to the finite photon conversion distance) with no decay electron
tags~\cite{ncpi0_sb}.  The tracks can be contained within SciBar or
penetrate into the EC.  Particle identification cuts based on energy
deposition require electron-like tracks in SciBar, and if the
tracks penetrate into the EC the energy deposit must be
shower-like---minimum ionizing particles are rejected.  The two tracks
are required to point to a common vertex within SciBar to eliminate
backgrounds from neutrino interactions upstream of SciBar.
Figure~\ref{fig:ncpi0_sb} shows the reconstructed pion mass
distribution for events contained within SciBar. The analysis is
ongoing.


\section{Antineutrino Cross Section Measurements}
\label{sec:nubar}

The search for charge-parity (CP) symmetry violation in the lepton
sector is one of the ultimate goals of the worldwide neutrino
programme.  If the effect is large enough, it could be observed via a
difference in the rates of $\nu_{\mu}\rightarrow\nu_e$ and
$\overline{\nu}_{\mu}\rightarrow\overline{\nu}_e$ oscillations.
However, even the most optimistic scenarios predict that the
difference will be relatively small, so the uncertainties on those
measurements must also be small.  This realisation motivates the need
for improved understanding of antineutrino-nucleus cross section
measurements.

\begin{figure}
\center
{\includegraphics[width=2.3in]{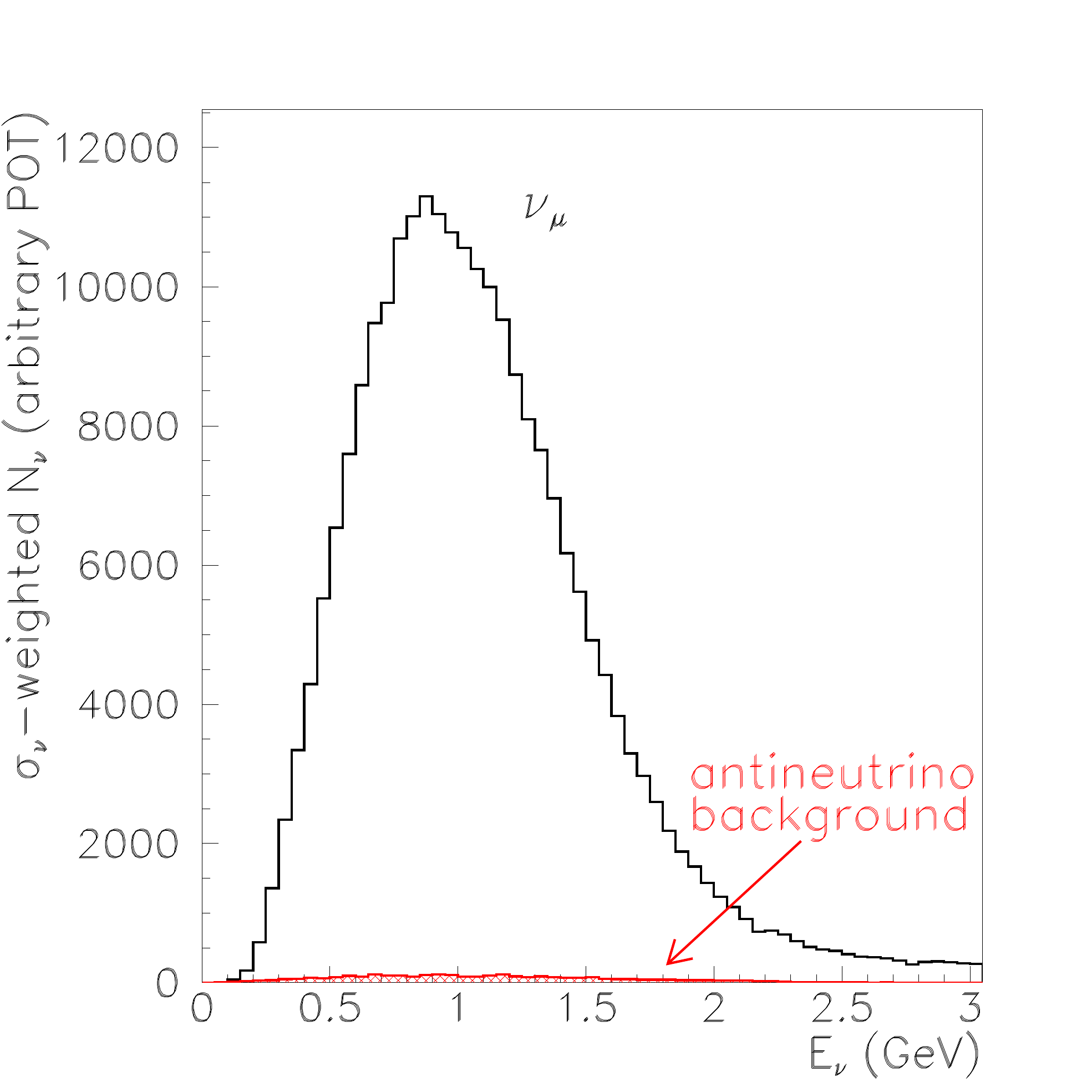}
\includegraphics[width=2.3in]{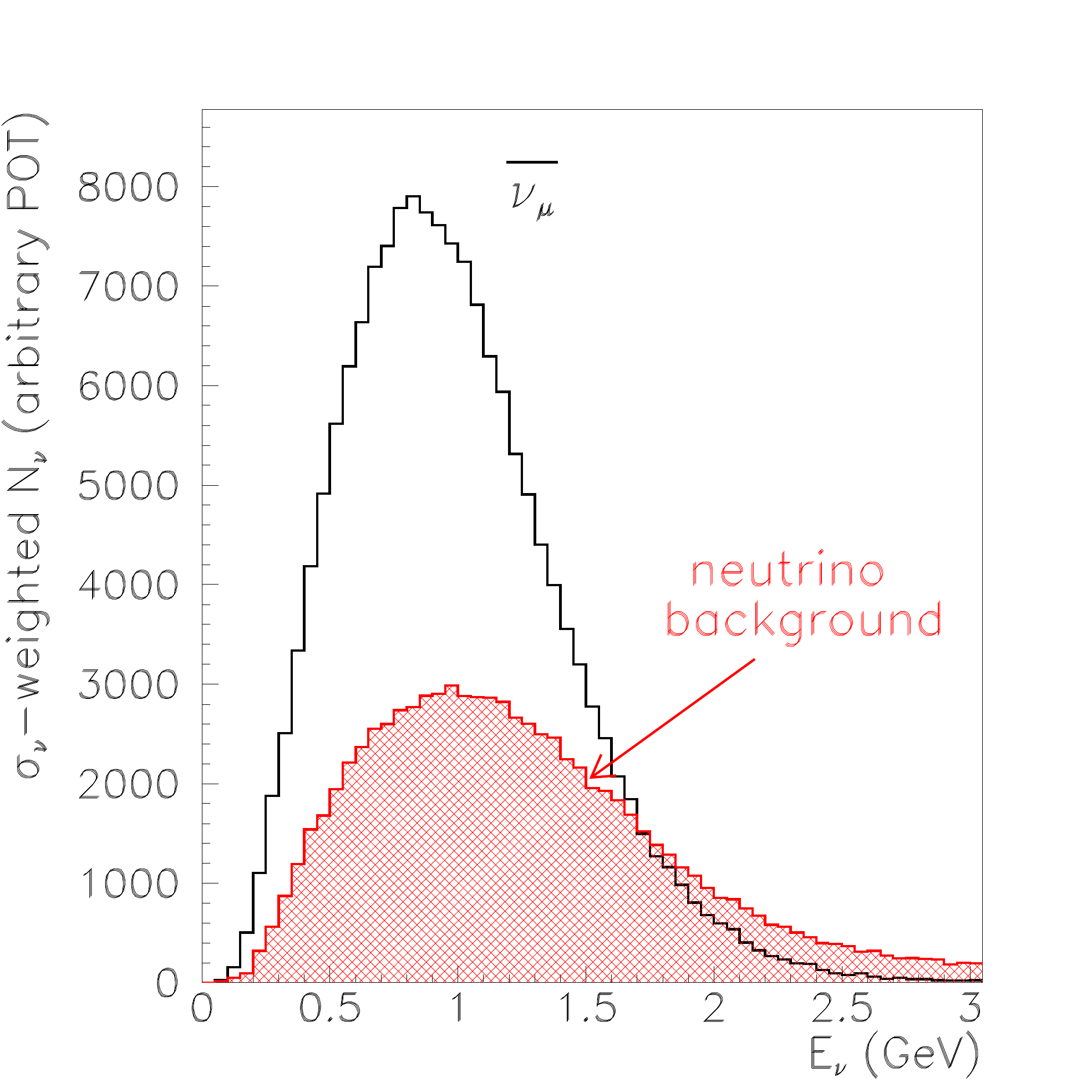}}
\caption{\em Monte Carlo simulation of cross section weighted events
  in the MiniBooNE detector, showing right sign (black histogram) and
  wrong sign (red cross-hatched histogram) events. Neutrino mode
  (left) and antineutrino mode (right) spectra are
  shown. \cite{mb_loi}}
\label{fig:nubar_ws}
\end{figure}

Figure~\ref{fig:past} illustrates the paucity of antineutrino cross
section measurements at low energy---there are no measurements of any
process below 1~GeV.  Antineutrino-nucleus cross sections are
significantly smaller---about 50\%---than neutrino-nucleus
cross sections. Moreover, $\pi^+$ production in neutrino targets,
leading to neutrino flux, is about twice as high as $\pi^-$
production, leading to antineutrino flux.  Together these reductions
cause an event rate in antineutrinos that is roughly 25\% of the
neutrino event rate per proton on target.

\subsection{Wrong Sign Backgrounds}

Most accelerator neutrino beams use magnetic focussing horns to
increase their neutrino fluxes; reversing the horn's current allows
selection of the oppositely charged mesons creating an antineutrino
beam.  However, paths within the inner conducter of cylindrical horns
do not cross any magnetic field lines, so mesons propagating down the
center are unaffected.  Combined with the fact that $\pi^+$ production
is far greater than $\pi^-$ production at the relevant energies, a
significant contamination of neutrinos is found in accelerator 
antineutrino beams.  Because they are produced by oppositely charged
parent mesons (and will produce oppositely charged leptons in CC
interactions) we call these wrong-sign (WS) backgrounds.  The
magnitude of wrong sign backgrounds in the BNB is illustrated in
figure~\ref{fig:nubar_ws}~\cite{sam,mb_loi}.

When studying the potential for antineutrino running, MiniBooNE
developed several techniques to mitigate WS
backgrounds~\cite{mb_loi,nuint05_nubar}.  These novel techniques are
necessary because (current) open volume Cherenkov detectors do not
have magnetic fields, so sign selection on an event-by-event basis is
impossible.  MiniBooNE's WS analysis techniques allow extraction of
the total fraction of WS events, but not as a function of energy.
MiniBooNE collected antineutrino data from January 2006 until Aug
2007, and then began collecting antineutrino again in April 2008.

\begin{figure}
\center
{\includegraphics*[width=2.4in,viewport=10 200 195 370]{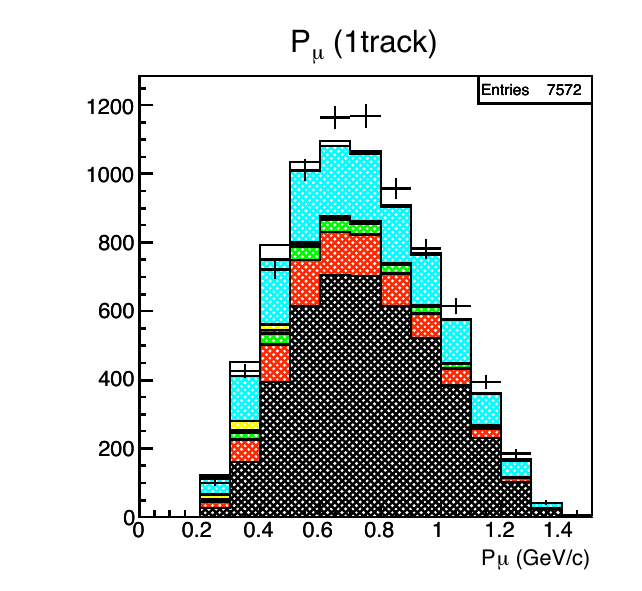}
\includegraphics*[width=2.4in,viewport=10 200 195 370]{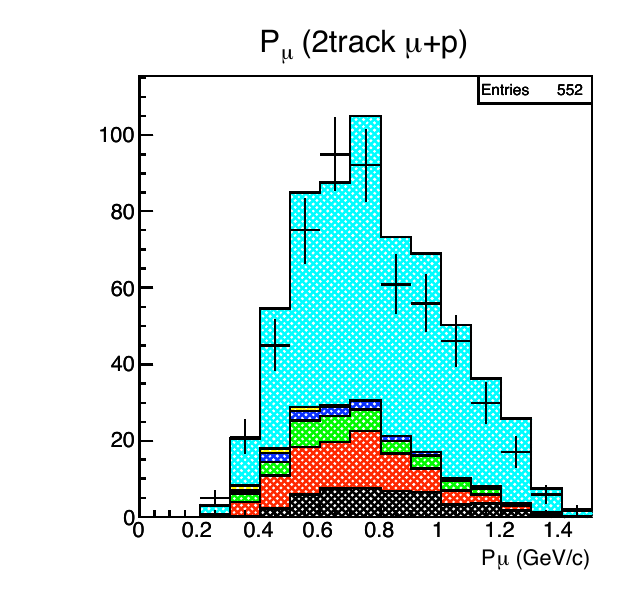}
\caption{\em SciBooNE antineutrino right-sign (left) and wrong-sign
  (right) muon momentum distributions. The data (black points) are
  shown with statistical uncertainties and MC wrong-sign (cyan) and
  right-sign (all other colours) contributions are
  indicated.\cite{nubarcccohpi_sb}.}
\label{fig:wsbg_sb}}
\end{figure}

SciBooNE's fine-grained vertex resolution offers a technique not
available to MiniBooNE.  Whereas neutrino \ccqe events, \numuccqe,
have two charged particles emerging from the neutrino interaction
vertex, antineutrino events, \numubarccqe, have just one.  So, by
simply selecting one-track events, SciBooNE creates a data sample with
80\% right sign events; by selecting two-track \mup events a 70\% WS
sample is obtained.  The results are shown in
figure~\ref{fig:wsbg_sb}~\cite{nubarcccohpi_sb}.

\subsection{SciBooNE \numubar CC Coherent Pion Search}

Charged-current coherent pion production forms a small fraction of
\ccpip events, making the coherent pion search essentially an exercise
in reducing and constraining backgrounds.  Interestingly, most
coherent pion production models (including the RS model used by most
neutrino generator MCs and experiments) predict that coherent pion
production by antineutrinos should have about the same cross section
as production by neutrinos.  Thus we can expect that antineutrino
searches should be more sensitive since the background rates, per
proton on target, will be reduced relative to neutrino searches.

\begin{figure}
\center
{\includegraphics*[width=4.0in,height=2.0in,viewport=570 200 720 350]{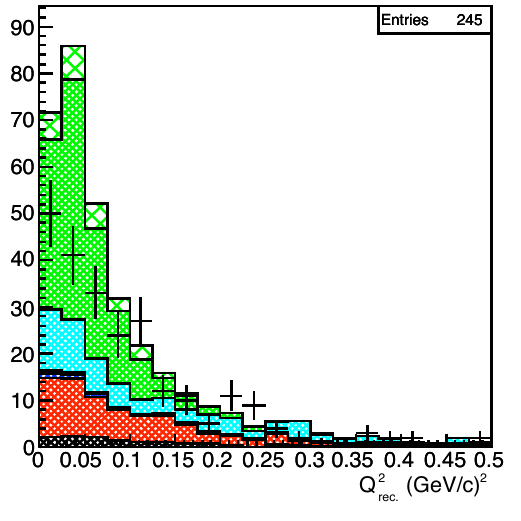}
\caption{\em SciBooNE antineutrino coherent pion sample reconstructed
  \q2 \, distribution. Data (black crosses) are shown with statistical
  uncertainty, as well as  MC CC coherent pion (green), wrong-sign (cyan),
  \numubar \ccqe (red) and \numubar \ccpip (incoherent) contributions
  from the MC.~\cite{nubarcccohpi_sb}.}
\label{fig:nubarcccohpi_sb_q2}}
\end{figure}

SciBooNE collected antineutrino data from June 2007 until August 2007,
and then again from April until August 2008. The SciBooNE antineutrino
CC coherent pion search follows the same chain of analysis events as
the neutrino search~\cite{nubarcccohpi_sb}.  The coherent pion
enriched sample is comprised of two-track, \mupi, low vertex activity,
non-QE events with forward going pions.  Then the CC coherent sample
is selected by requiring that the reconstructed \q2 \, (assuming a
\ccqe hypothesis) be less than 0.1~\gev2 .  The SciBooNE antineutrino
CC coherent pion data and MC are shown in
figure~\ref{fig:nubarcccohpi_sb_q2}.

The data clearly lie above the MC predicted backgrounds in the
coherent pion region below 0.1~\gev2 , but the predicted coherent pion
signal is larger than what is observed in the data.  These preliminary
SciBooNE data suggest non-zero coherent pion production, but it
appears to be lower than the level predicted by the RS model employed
by the NEUT generator. It is interesting to note that if the data
excess above the predicted backgrounds were converted into a cross
section ratio, it would be consistent with the SciBooNE (and K2K)
upper limits observed in the neutrino mode
search~\cite{nubarcccohpi_sb}.  Studies are ongoing.


\section{Summary and Conclusions}
\label{sec:summary}

The physics of neutrino-nucleus interactions near 1~GeV is today a
vibrant field being driven by multiple experiments across the world
making new measurements.  Although the field lay dormant for nearly 20
years, the recent generation of accelerator neutrino oscillation
experiments has revived it with an injection of new, high statistics
and high quality data.  Recent data from K2K, MiniBooNE and SciBooNE
have revealed multiple discrepancies between the industry standard
models and new observations.  We have summarised the best-studied such
discrepancies---those that weigh the heaviest on neutrino oscillation
searches.  At present there is no clear prescription on which of many
possible theory paths forward will yield the best results; in other
words there are ample opportunities.  By way of summarising, we pose
the open questions revealed by the data presented herein.

Why do recent \ccqe data near 1~GeV suggest a high value of \maqe ,
and what does such a high value mean?  We did not cover it, but at
higher energies (3-100~GeV) the NOMAD experiment observes a value of
\maqe = 1.05$\pm$0.02(stat)$\pm$0.06(syst)~GeV/c$^2$~\cite{nomad}.
How is this reconciled with the observations from K2K and MiniBooNE?

What is the source of the low \q2 \, deficit?  This phenomenon has been
observed in both \ccqe and CC non-QE channels.  MiniBooNE addressed
the problem by introducing a scaling parameter for Pauli blocking
within the context of the RFG model. Colloquially, we can say that
MiniBooNE found the RFG model didn't have enough knobs to turn so they
added one. Why not? Recent work suggests that in the \ccpip channels
the low \q2 \, issue can be addressed by careful adjustment of the form
factors within the RS model~\cite{jarek_nuint09}. Does the resolution
of this issue require better nuclear modeling as well?

How do we reconcile the apparently disparate measurements of
coherent pion production in CC and NC channels?  Most theoretical
models agree on the close relationship between NC and CC coherent pion
production but SciBooNE and K2K have set strict limits on the relative
amount of CC coherent pion production while MiniBooNE has shown clear
evidence for NC coherent pion production.  Can the hint that SciBooNE
has seen in antineutrino data resolve the issue?

How can we improve the nuclear model used for neutrino scattering?
Can we converge on a uniform treatment of final state interactions?
Electron scattering experiments have shown conclusively that there are
strong intra-nuclear correlations~\cite{nuke_correlations}, but the
RFG assumes none. Is the low \q2 \, issue just a matter of FSI? Very
interesting lectures were given in L\k{a}dek on these
topics~\cite{dytman,benhar}.

Because of the complexity of testing new models against published
data, a new concensus is emerging that we experimentalists must strive
to publish POT-normalised differential cross sections of final state
particles.  We believe this is the key to reconciling the wealth of
new data being collected with the many new theory ideas published in
recent years---and we hope that many more new ideas will be inspired
by the new data.

May we continue to live in interesting times!


\section{ACKNOWLEDGMENTS}
\label{sec:ack}

The author would like to sincerely thank the organisers, especially
J. Sobczyk, for the invitation to the L\k{a}dek school, which was a
very stimulating intellectual environment.  Dzi\k{e}kuj\k{e}!  Many
thanks also go to G.~P. Zeller and J.~Monroe for many hours of
fruitful discussion in the preparation of the lectures and these
proceedings.


\end{document}